\documentclass[emulateapj,twocolumn]{aastex63}

\usepackage{amsmath}
\usepackage{natbib}
\usepackage{mathtools}
\usepackage{float}
\usepackage{bm}
\restylefloat{table}

\newcommand{\be}{\begin{equation}} 
\newcommand{\ee}{\end{equation}}

\newcommand{\ba}{\begin{eqnarray}} 
\newcommand{\ea}{\end{eqnarray}}

\newcommand{\stheta}{s_{\theta}}
\newcommand{\ctheta}{c_{\theta}}
\newcommand{\wtide}{\omega_{mm'}}
\newcommand{\wtilde}{\tilde{\omega}_{mm'}}
\newcommand{\hats}{\bm {\hat s}}
\newcommand{\hatl}{\bm {\hat l}}
\newcommand{\hatz}{\bm {\hat z}}
\newcommand{\hatx}{\bm {\hat x}}
\newcommand{\haty}{\bm {\hat y}}
\newcommand{\vecs}{\bm {S}}
\newcommand{\vecl}{\bm {L}}
\newcommand{\D}{{\rm d}}
\newcommand{\Q}{Q_{m m'}'}
\newcommand{\mplanet}{M_{\rm p}}
\newcommand{\rplanet}{R_{\rm p}}
\newcommand{\rtide}{R_{\rm tide}}
\newcommand{\teff}{T_{\rm eff}}
\newcommand{\feh}{[\rm Fe/H]}

\newcommand{\msun}{M_\odot}
\newcommand{\rsun}{R_\odot}

\newcommand{\mjup}{M_{\rm J}}
\newcommand{\rjup}{R_{\rm J}}

\shorttitle{Tidal Realignment in Hot Jupiter Systems}
\shortauthors{K. R. Anderson et al.}

\begin{document}

\title{On a Possible Solution to the Tidal Realignment Problem for Hot Jupiters}

\correspondingauthor{Kassandra Anderson}
\email{kassandra@princeton.edu}
\author[0000-0002-7388-0163]{Kassandra R.\ Anderson} 
\altaffiliation{Lyman Spitzer, Jr.\ Postdoctoral Fellow}
\affiliation{Department of Astrophysical Sciences, Princeton University, Princeton, NJ 08544, USA}

\author[0000-0002-4265-047X]{Joshua N.\ Winn}
\affiliation{Department of Astrophysical Sciences, Princeton University, Princeton, NJ 08544, USA}

\author[0000-0003-4464-1371]{Kaloyan Penev} 
\affiliation{Department of Physics, The University of Texas at Dallas, 800 West Campbell Road, Richardson, TX 75080-3021, USA}

\begin{abstract}
Hot stars with hot Jupiters have a wide range of obliquities, while cool stars with hot Jupiters tend to have low obliquities. An enticing explanation for this pattern is tidal realignment of the cool host stars, although this explanation assumes that obliquity damping occurs faster than orbital decay, an assumption that needs
further exploration. Here we revisit this tidal realignment problem, building on previous work identifying a low-frequency component of the time-variable tidal potential that affects the obliquity but not the orbital separation.  We adopt a recent empirically-based model for the stellar tidal quality factor and its sharp increase with forcing frequency. This leads to enhanced dissipation at low frequencies, and efficient obliquity damping.  We model the tidal evolution of 46 observed hot Jupiters orbiting cool stars. A key parameter is the stellar age, which we determine in a homogeneous manner for the sample, taking advantage of Gaia DR2 data. We explore a variety of tidal histories and futures for each system, finding
in most cases that the stellar obliquity is successfully damped before the planet is destroyed. A testable prediction of our model is that hot-Jupiter hosts with orbital periods shorter than 2--3 days should have obliquities much smaller than $1^\circ$. With the possible exception of WASP-19b, the predicted future lifetimes of the planets range from $10^8$\,yr to more than $10^{10}$\,yr. Thus, our model implies that these hot Jupiters are probably not in immediate danger of being devoured by their host stars while they
are on the main sequence.
\end{abstract}


\section{Introduction}

Stellar obliquities (spin-orbit angles) are interesting aspects of the
architectures of exoplanetary systems because they may shed light on the formation and migration processes that lead to close-orbiting planets. Proposed migration theories for the existence of hot Jupiters include migration due to disk torques \citep[e.g.][]{lin1996,kley2012}, or some form of high-eccentricity migration, in which the planet's eccentricity becomes excited and the periastron distance becomes small enough for tides to shrink and circularize the orbit \citep[e.g.][]{wu2003,fabrycky2007,wu2011,beauge2012}. In recent years, the possibility for {\it in situ} formation has been considered \citep{boley2016,batygin2016}. See also \cite{dawson2018} for a recent review. High-eccentricity migration naturally produces short-period planets on misaligned orbits relative to the host star. Disk migration or {\it in situ} formation may also result in large stellar spin-orbit misalignments if the protoplanetary disk itself was initially misaligned \citep[e.g.][]{bate2010,lai2011,batygin2012}.

More than a decade of observations of stars with close-orbiting giant
planets has revealed a wide range of obliquities, including closely aligned systems, nearly perpendicular systems, and retrograde systems \citep{WinnFabrycky2015,Triaud2018}.  The observations have also revealed an
intriguing pattern, first noted by \cite{winn2010} and \cite{Schlaufman2010}.  Stars with effective temperatures
between about 6200 and 7000\,K show a very broad range of
obliquities.  In contrast, stars with effective temperatures between
about 5000 and 6200\,K usually have low obliquities, except
for a few cases of relatively low-mass planets ($\lesssim$0.4\,$M_{\rm Jup}$) on
relatively wide orbits ($a/R_\star \gtrsim 6$).

An enticing explanation for these trends is tidal realignment.  In this scenario,
hot Jupiters begin with a wide range of orbital orientations, and tidal torques
gradually align the orbit and the star.  However, the rate of tidal alignment depends on the orbital separation, planet-to-star mass ratio, and the structure of the star.
The only systems for which alignment is rapid enough to align the star
and orbit on an astrophysically relevant timescale are those with massive planets,
close orbits, and stars with large surface convective zones (which are
thought to enhance the rate of tidal dissipation).  This explanation is corroborated by the finding that the transition from low obliquities to a broad range of
obliquities occurs at an effective temperature of about 6200\,K, which is
roughly the temperature beyond which the extent of the surface convective region rapidly diminishes. 

Previous works involving stellar tides in hot Jupiter systems have focused on a variety of issues, such as planet destruction and stellar spin-up \citep[e.g.][]{jackson2009,teitler2014}, tidal equilibrium states \citep[e.g.][]{lanza2009,damiani2015}, and the effects of stellar evolution \citep[e.g.][]{penev2014,bolmont2016}. Various works have considered the tidal dynamics of spin-orbit misaligned systems and the tidal realignment hypothesis \citep[e.g.][]{barker2009,matsumura2010,Valsecchi+2014}. A problem
with this hypothesis is that simple tidal theories and standard assumptions lead to the conclusion that obliquity damping occurs on the same timescale as the shrinkage of the planetary orbit -- implying that planets would typically be destroyed during the process of tidal realignment.

A solution to this problem was proposed by \cite{lai2012}, who pointed out that when
the tidal potential of a spin-orbit misaligned system is expanded as a Fourier series, the torques that arise from some of the components affect the obliquity but not the orbital distance. More precisely, in a spin-orbit misaligned system, the possible
frequencies of the tidal perturbations are linear combinations of the star's spin frequency $\Omega_\star$ and the planet's orbital frequency $\Omega$. The components with frequencies that involve only $\Omega_\star$ and not $\Omega$ allow angular momentum to be transferred between the spin and orbit, but not energy. Therefore, energy must be dissipated at the expense of the stellar spin, rather than by changing the orbital distance.  The obliquity-only components could excite
inertial waves in the convective envelope of the star, the damping of which would enhance the rate of dissipation \citep[see, e.g.][]{ogilvie2007,goodman2009}. Since inertial waves are only excited for tidal frequencies less than twice the stellar rotation frequency, inertial waves typically affect only the obliquity in hot Jupiter systems, and not the orbital distance. A study of tidal dissipation in spin-orbit misaligned systems due to the excitation of inertial waves in convective envelopes was performed by \cite{lin2017}, who found efficient dissipation for the obliquity-only components. Although various uncertainties remain, dissipation of inertial waves in convective envelopes thus seems to be a promising theoretical mechanism for damping obliquities rapidly compared to orbital decay, provided that the tidal components that lead to orbital decay have frequencies outside of the inertial wave regime.

Following the suggestion made by \cite{lai2012}, subsequent studies of tidal evolution in spin-orbit misaligned systems have further explored this issue. \cite{rogers2013} considered in isolation the main tidal component that only affects the obliquity, and found obliquity realignment to be possible only when the system is initially prograde. Initially retrograde systems were driven towards either $90^\circ$ or $180^\circ$. This seemed inconsistent with observations, which do not appear to show an excess of systems with obliquities near $90^\circ$ or $180^\circ$. However, considering the contributions from all the tidal components, 
\cite{xue2014} found that occupancy of the $90^\circ$ and $180^\circ$ states is only
temporary -- obliquities eventually damp to $0^\circ$. \cite{li2016} included the
effects of magnetic braking, and came to the same
conclusion. The tidal realignment hypothesis thus remains a viable explanation for at least some aspects of the obliquity-temperature trend.

In this paper, we investigate a variation of this solution to the tidal realignment problem, utilizing the latest empirical estimates of the modified stellar tidal quality factor ($Q'$) in hot Jupiter systems. Based on the anomalously rapid rotation that has been observed for some hot Jupiter hosts, and observed trends in the amount of excess rotation as a function of system age, orbital period, and other
properties, \cite{penev2018} found evidence that $Q'$ increases strongly with the tidal forcing frequency.  For simplicity, their calculations assumed 
spin-orbit alignment throughout the evolution.
For circular, spin-orbit aligned systems only a single component of the Fourier expansion of the tidal potential is non-zero, and operates at a frequency $2(\Omega - \Omega_\star)$. If such scaling with forcing frequency extends to the other tidal components for spin-orbit misaligned systems (so that each component has a different rate of dissipation), the enhanced dissipation at the lowest forcing frequencies may lead to obliquity realignment before the planetary orbit catastrophically decays. \cite{penev2018} suggested that a frequency-dependent quality factor could help to solve the tidal realignment problem for hot Jupiters, but left detailed exploration for future work.

This paper reports on our effort to follow up on the suggestion
by \cite{penev2018} and test whether it is a viable solution to
the tidal realignment problem.
We conducted an in-depth numerical study of the tidal evolution of spin-orbit misaligned hot Jupiter systems, implementing a frequency-dependent $Q'$ for each distinct tidal component, and accounting for stellar magnetic braking. We selected a sample of hot Jupiters orbiting cool stars and determined the basic stellar
properties (including the age) by fitting stellar-evolutionary models
to spectroscopic, photometric, and Gaia DR2 parallax measurements.
We then considered possible tidal histories for each planet, representing
different formation and migration channels, and evaluated the prospects for efficient obliquity realignment.

This paper is organized as follows. In Section \ref{sec:model}, we present the details of the tidal model along with illustrative examples of successful obliquity realignment prior to planet destruction. In Section \ref{sec:parameter}, we use
synthetic populations of planets to explore how the results depend upon various model parameters, and compare the results with the overall properties of observed systems. In Section \ref{sec:observed}, we specialize to a sample of 46 real hot Jupiter
systems. We describe our homogeneous analysis of the stellar parameters, and then explore a wide range of possible tidal histories and futures for each planet. We conclude in Section \ref{sec:conclusion}.

\section{Tidal and Magnetic Braking Model} \label{sec:model}
\subsection{Setup and Governing Equations}
We consider a star-planet system with masses and radii $M_\star$, $R_\star$ and $\mplanet$, $\rplanet$. We assume the star is rotating uniformly with frequency $\Omega_\star$, and has a moment of inertia $I_\star = k_\star M_\star R_\star^2$ with $k_\star \simeq 0.06$. The stellar angular momentum is denoted $\vecs = S \hats$ and the orbital angular momentum as $\vecl = L \hatl$, with $\hats$ and $\hatl$ unit vectors. We restrict our attention to circular, possibly inclined orbits, having in mind that the planet's orbital
eccentricity has already been extensively damped and
the planet's rotation is synchronized with its orbit.
We consider the tides raised by the planet on the host star, in concert with the magnetic braking torque on the host star.

The spin and orbital angular momenta each evolve due to the tidal torque and energy transfer. We adopt the tidal model of \cite{lai2012}, which parameterizes the dissipation in terms of a separate phase-lag for each tidal component. Below, we summarize the key features of this model.

Placing $\hats$ along the $\hatz$-axis and $\hatl$ in the $x-z$ plane, the as-yet unspecified tidal torque and energy dissipation rate are ${\bf T} =  T_x \hatx + T_y \haty + T_z \hatz$ and $\dot{E}$ respectively. The $\hatx$ and $\hatz$ components introduce dissipation, leading to changes in orbital distance, spin rate, and obliquity. The $\haty$ component contributes only to precession, and does not
otherwise affect the orbital and spin parameters; for
this reason, we did not include the $\haty$ component in
our calculations. The spin and orbit thus evolve under tides as
\ba
\frac{\D \vecs}{\D t} & = & T_x \hatx + T_z \hatz \nonumber \\
\frac{\D \vecl}{\D t} & = & - \frac{\D \vecs}{\D t},
\ea
In terms of semi-major axis $a$, spin frequency $\Omega_\star$, and obliquity $\theta \equiv \cos^{-1}(\hats\cdot\hatl)$, the evolution equations take the form
\ba
\bigg( \frac{\D a}{\D t} \bigg)_{\rm Tide} & = & - \frac{2 a^2 \dot{E}}{G M_\star \mplanet}, \label{eq:adot} \\
\bigg( \frac{\D \Omega_\star}{\D t} \bigg)_{\rm Tide} & = & \frac{T_z}{I_\star}, \label{eq:Omegadot} \\ 
\bigg( \frac{\D \theta}{\D t} \bigg)_{\rm Tide} & = & -\frac{T_x}{S} - \frac{T_x \cos \theta}{L} + \frac{T_z \sin \theta}{L}. \label{eq:thetadot}
\ea

The tidal torque components and energy dissipation rate are obtained by expanding the tidal potential in spherical harmonics and transforming to the rotating frame of the host star to analyze the fluid response (see Sections 2.1--2.3 of \cite{lai2012}). For a circular, inclined orbit, there are seven independent Fourier components of the tidal potential, operating at frequencies
\be
\wtide = m' \Omega - m \Omega_\star.
\ee
For a derivation of the tidal components for circular, spin-orbit misaligned systems, see \cite{barker2009}. The physically distinct tidal components that lead to dissipation are $(m,m') = (0,2)$, $(\pm 1,2)$, $(\pm 2,2)$, $(1,0)$, and $(2,0)$.  For each component, the dissipation is specified by a phase shift $\Delta_{mm'}$ and a Love coefficient $\kappa_{mm'}$. Since only the product of
these two parameters is relevant to our calculations, we define $\Delta'_{mm'} \equiv \Delta_{mm'} \kappa_{mm'}$. The torque components $T_x$ and $T_z$ and the energy dissipation rate $\dot{E}$ are calculated as a sum over contributions from all the tidal components, and can be written explicitly as (see \citealt{lai2012}, equations 27, 28, 35)
\begin{flalign}
T_x & = \frac{3 \pi}{20} T_0 \bigg[\frac{1}{2} \stheta (1 + \ctheta)^3 \Delta'_{22} 
 + \stheta (1 + \ctheta)^2 (2 - \ctheta) \Delta'_{12} & \nonumber \\
 &+ 3 \stheta^3 \Delta'_{02}
 + \stheta (1 - \ctheta)^2 (2 + \ctheta) \Delta'_{-12} 
 + \frac{1}{2} \stheta (1 - \ctheta)^3 \Delta'_{-22} \bigg] & \nonumber \\
 & - \frac{3 \pi}{5} T_0 \bigg[\frac{1}{2} \stheta^3 \ctheta \Delta'_{20} + \stheta \ctheta^3 \Delta'_{10} \bigg],
\end{flalign}
\begin{flalign}
T_z & = \frac{3 \pi}{20} T_0 \bigg[ \frac{1}{2} (1 + \ctheta)^4 \Delta'_{22}
+ \stheta^2(1 + \ctheta)^2 \Delta'_{12} & \nonumber \\
& -  \stheta^2 (1 - \ctheta)^2 \Delta'_{-12} 
- \frac{1}{2} (1 - \ctheta)^4 \Delta'_{-22} \bigg] & \nonumber \\
& + \frac{3 \pi}{5} T_0 \bigg[\frac{1}{2} \stheta^4 \Delta'_{20} + \stheta^2 \ctheta^2 \Delta'_{10}  \bigg],
\end{flalign}
\begin{flalign}
\dot{E} & = \frac{3 \pi}{20} T_0 \Omega \bigg[ \frac{1}{2}(1 + \ctheta)^4 \Delta'_{22} 
+ 2 \stheta^2 (1 + \ctheta)^2 \Delta'_{12} & \nonumber \\
& + 3 \stheta^4 \Delta'_{02}
  + 2 \stheta^2 (1 - \ctheta)^2 \Delta'_{-12}
  + \frac{1}{2} (1 - \ctheta)^4 \Delta'_{-22} \bigg],
\end{flalign}
where $\ctheta = \cos \theta$, $\stheta = \sin \theta$, and
\be
T_0 = \frac{G \mplanet^2 R_\star^5}{a^6}.
\ee
This tidal model thus allows for general specification of each phase lag $\Delta'_{mm'}$, which we relate to the modified quality factor $Q'_{mm'}$ via $Q'_{mm'} = 1 / |\Delta'_{mm'}|$\footnote{Note that the literature
contains alternative definitions relating quality factor to phase lag, which differ by numerical coefficients of order unity.}. 

By modeling the tidal spin-up of hot Jupiter hosts and
comparing to observations,
\cite{penev2018} found evidence that
the $m = m' = 2$ component of the stellar modified tidal quality factor exhibits the frequency dependence
\be
Q_{22}' \propto \omega_{22}^3.
\label{eq:Q}
\ee
We decided to investigate a more
general power-law dependence, $\Q \propto \omega_{mm'}^{p}$, with a single value of $p$ that applies to all
the tidal components. We thus adopt phase shifts of the form
\be
\Delta'_{mm'} \propto \wtide^{-p}.
\label{eq:phaselag}
\ee
Most of the results we will describe
are for the case $p = 3$, motivated by the work
of \cite{penev2018}, but we also investigated some other possible values of $p$. We note that $p = -1$  corresponds to a constant time lag, while $p = 0$ corresponds to a constant phase lag.

As the frequency decreases, the scaling in equation (\ref{eq:phaselag}) must
break down at some point, and when the tidal frequency approaches zero, the dissipation should vanish. We therefore introduce a softening parameter $\epsilon$ and impose a maximum (in magnitude) phase shift $\Delta'_{\rm max}$, so that the phase shift for the $mm'$ component takes the form
\be
\Delta'_{mm'} = {\rm sign} (\wtilde) \min \left[ \frac{\Delta'_0 | \wtilde |}{(|\wtilde| + \epsilon)^{p+1}},\Delta'_{\rm max} \right],
\label{eq:toyphaselag}
\ee
where
\be
\wtilde \equiv \frac{\wtide}{2 \pi \ {\rm days}^{-1}},
\ee
and where the scaling constants $\Delta'_0, \Delta'_{\rm max}$ are positive quantities. For $|\wtide| \gg \epsilon$, equation (\ref{eq:toyphaselag}) yields $\Delta'_{mm'} \propto \wtide^{-p}$, and for $|\wtide| \ll \epsilon$, $\Delta'_{mm'} \to 0$. The scaling found by \cite{penev2018} was calibrated using hot Jupiter systems with tidal forcing periods  as short as 0.5 days. This implies that the softening parameter should satisfy $\epsilon \ll 1$. Somewhat arbitrarily,
we chose $\epsilon = 0.05$. We experimented with values of $\epsilon$ a factor of 2 larger and smaller than 0.05, and found no qualitative differences in the results of individual integrations, and no difference at all in the statistical results.

\begin{figure}
\centering 
\includegraphics[scale=0.5]{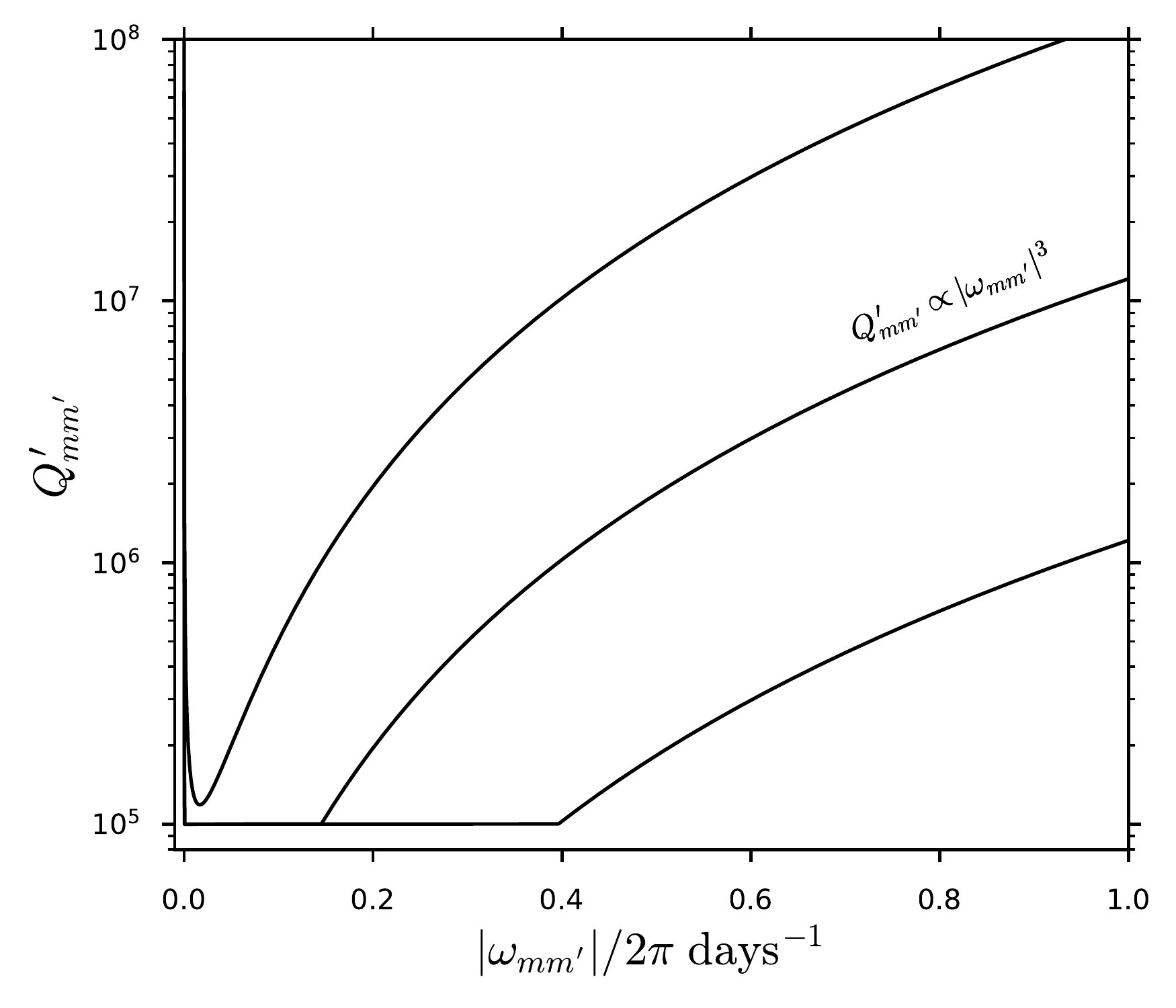}
\caption{Frequency dependence for the stellar tidal quality factors adopted in this paper $\Q = 1/|\Delta'_{mm'}|$, using equation (\ref{eq:toyphaselag}), and setting $p = 3$, $Q'_{\rm min} = 1 / \Delta'_{\rm max} = 10^5$, and $\epsilon = 0.05$. The curves denote (from top to bottom), $Q'_0 = 1/\Delta'_0 = 10^8, 10^7, 10^6$.}
\label{fig:Q}
\end{figure}

With these choices, there is a minimum possible
value for the modified quality factor $Q'_{\rm min} = 1/\Delta'_{\rm max}$ and an overall scale $Q'_0 = 1/\Delta'_0$. We chose $Q'_{\rm min} = 10^5$, motivated by observations of eclipsing binaries \citep{milliman2014}. The adopted scaling of $Q_{mm'}$ (from equation \ref{eq:toyphaselag}) is depicted in Fig.~\ref{fig:Q}. The results of \cite{penev2018} indicate that $Q'_0 \sim 10^6$. In Section \ref{sec:parameter} we explore the implications of varying $Q_{0}'$, and in Section \ref{sec:observed} we adjust the exact value of $Q'_0$ system-by-system in order to reproduce the observed spin and orbital periods.

Note that in this model, $Q'_{mm'}$ loses its frequency dependence and becomes a constant for low
enough tidal frequencies:
\be
| \wtide | \leq \frac{2 \pi}{\rm days} \left( \frac{Q'_{\rm min}}{Q'_0} \right)^{1/p}.
\ee
For the $(1,0)$ component, this corresponds to a spin frequency 
\be
\Omega_\star \leq \frac{2 \pi}{\rm days} \left( \frac{Q'_{\rm min}}{Q'_0} \right)^{1/p}.
\ee

Along with the tidal torques, we include the spin-down torque on the host star due to magnetic braking ($T_{\rm MB}$), via
\be
\bigg( \frac{\D \Omega_\star}{\D t} \bigg)_{\rm MB} = -\alpha \min (\Omega_{\rm sat}, \Omega_\star)^2\,\Omega_\star = -\frac{T_{\rm MB}}{I_\star},
\label{eq:mb}
\ee
\citep[see, e.g.,][]{skumanich1972,kawaler1988,gallet2013}, where $\Omega_{\rm sat}$ is the frequency at which the stellar dynamo saturates, and 
\be
\alpha \simeq 1.5 \times 10^{-14} {\rm yr}~ \bigg(\frac{M_\star}{\msun} \bigg)^{-3/2}  \bigg(\frac{R_\star}{\rsun} \bigg)^{-3/2}.
\label{eq:alpha}
\ee
The mass and radius scaling in equation (\ref{eq:alpha}) is motivated by equation (3) of \cite{amard2016}. The coefficient, taken from \cite{barker2009}, reproduces the spin period of the Sun at its current age. Although the dependence on stellar mass and radius is taken from spin-down models that may not be completely realistic, we will ultimately restrict our attention to a relatively narrow range of stellar effective temperatures to help alleviate this issue. We set $\Omega_{\rm sat} = 10\, \Omega_\odot$ throughout this paper.

In summary, the spin and orbit evolve under the influence of both tides and magnetic braking according to equations (\ref{eq:adot})-(\ref{eq:thetadot}) and (\ref{eq:mb}), using equation (\ref{eq:toyphaselag}) to specify the dissipation for each tidal component. In practice, we evolve the system in terms of the Cartesian components of the angular momentum vectors in an inertial frame, rather than in terms of orbital elements.

If at any point the orbital distance becomes less than the tidal disruption radius $\rtide$, with
\be
\rtide = \eta R_{\rm p} \left(\frac{M_\star}{M_{\rm p}} \right)^{1/3},
\label{eq:rtide}
\ee
we terminate the integration and consider the planet to be destroyed. The value of $\eta$ depends on the internal constitution of the planet, which is of course unknown.
Estimates based on numerical simulations are typically between 2 and 3 \citep[e.g.][]{guillochon2011}. Unless otherwise specified, we set $\eta = 2.5$ for the remainder of this paper. 

\subsection{General Features and Examples}
Here we discuss the general features of the tidal model, and present several illustrative examples of tidal evolution in spin-orbit misaligned systems.

As discussed by \cite{lai2012}, the $(m,m') = (1,0)$ and $(2,0)$ tidal components do not transfer energy, because they are static in the inertial frame. These components thereby realign obliquities without inducing orbital decay. For typical hot Jupiter systems, the orbital frequency $\Omega$ usually exceeds the spin frequency $\Omega_\star$ by a factor of at least a few.  Consequently, the $(m,m') = (1,0)$ component typically has the lowest forcing frequency of all
the tidal components. Under the assumption that $\Q \propto \wtide^p$ (with $p > 0$), $Q_{10}$ can be much smaller than the other $\Q$. This is why the model is
capable of
reducing the obliquity without destroying the planet.

It is well-known that the $(m,m') = (1,0)$ component alone may drive the obliquity towards $0^\circ$, $90^\circ$, or $180^\circ$ depending on the initial obliquity \citep{lai2012,rogers2013}. However, considering the combined effects of all the tidal components, the perpendicular and antiparallel obliquity states are at most temporarily stable \citep{xue2014,li2016}, although the system may reside in such states for long periods of time.

To illustrate this phenomenon using the specific tidal model adopted in this paper, we calculate $\dot{\theta}$ as a function of $\theta$, using equation (\ref{eq:thetadot}). We denote by $\theta^*$ the values of $\theta$ for which $\dot{\theta} = 0$ (i.e., the ``fixed points''\footnote{Strictly speaking, they are not true fixed points, because the orbital distance and spin period continue to evolve in these states.}). The sign of $\dot{\theta}$ in the neighborhood of $\theta^*$
is related to the stability of the fixed point.
We denote by $\theta^*_{-}$ and $\theta^*_{+}$ the values of $\theta$ just less than $\theta^*$, and just greater than $\theta^*$, respectively.
If $\dot{\theta} (\theta^*_-) > 0$ and $\dot{\theta} (\theta^*_+) < 0$, the system may be driven towards $\theta^*$.  In contrast, if $\dot{\theta} (\theta^*_-) < 0$ and $\dot{\theta} (\theta^*_+) > 0$ the system will be repelled from $\theta^*$. Finally, if the sign of $\dot{\theta}$ does not change from $\theta^*_-$ to $\theta^*_+$, $\theta^*$ is a saddle point; it is stable only when approaching from one direction. 

\begin{figure}
\centering 
\includegraphics[scale=0.53]{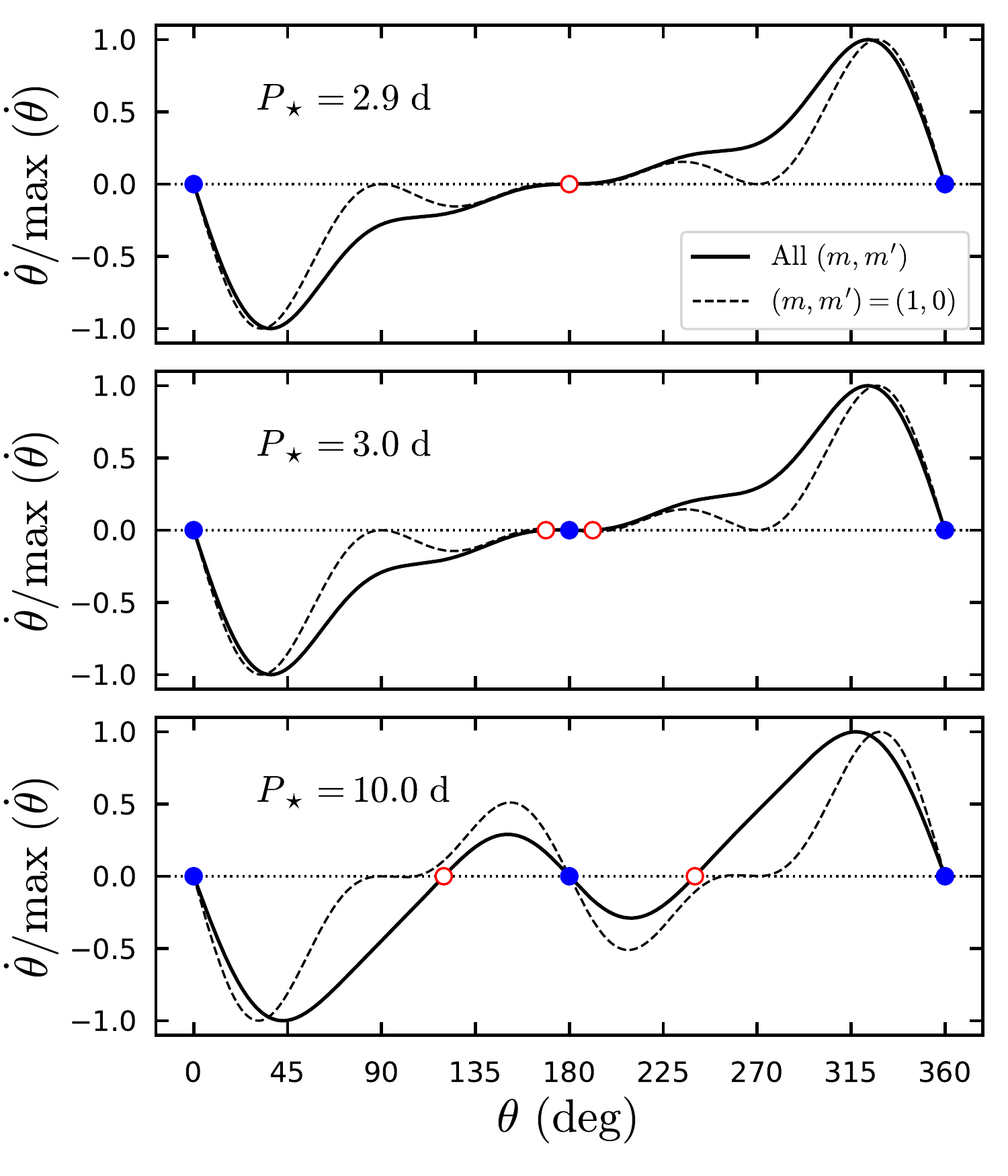}
\caption{Shown is $\dot{\theta}$ as a function of $\theta$ for a Jupiter-mass planet orbiting a solar-type host star. We have fixed the planet orbital period to $P_{\rm orb} = 2$ days, and varied the stellar spin period in each panel, as labeled. The tidal parameters are set to $Q_0' = 10^6$ and $p = 3$. The solid curve accounts for all values of $(m,m')$ that contribute to the torque, while the dashed curve shows the result including only the $(1,0)$ component. The blue (open red) circles indicate the stable (unstable) fixed points for the obliquity. The spin-orbit aligned state is stable, while the anti-aligned state may be stable or unstable depending upon the spin period. This indicates that the obliquity may spend periods of time locked at $180^\circ$.}
\label{fig:theta_dot_vs_theta}
\end{figure}

Figure \ref{fig:theta_dot_vs_theta} shows $\dot{\theta}$ as a function of $\theta$, with the tidal model parameters set to $p = 3$ and $Q_0' = 10^6$. We fixed the orbital period and vary the spin period. We compare results for the $(1,0)$ component in isolation, and the combined effects of all components. Although $90^\circ$ is a fixed point when considering the $(1,0)$ component alone, it is not a fixed
point when including all the other tidal components.

\begin{figure}
\centering 
\includegraphics[scale=0.55]{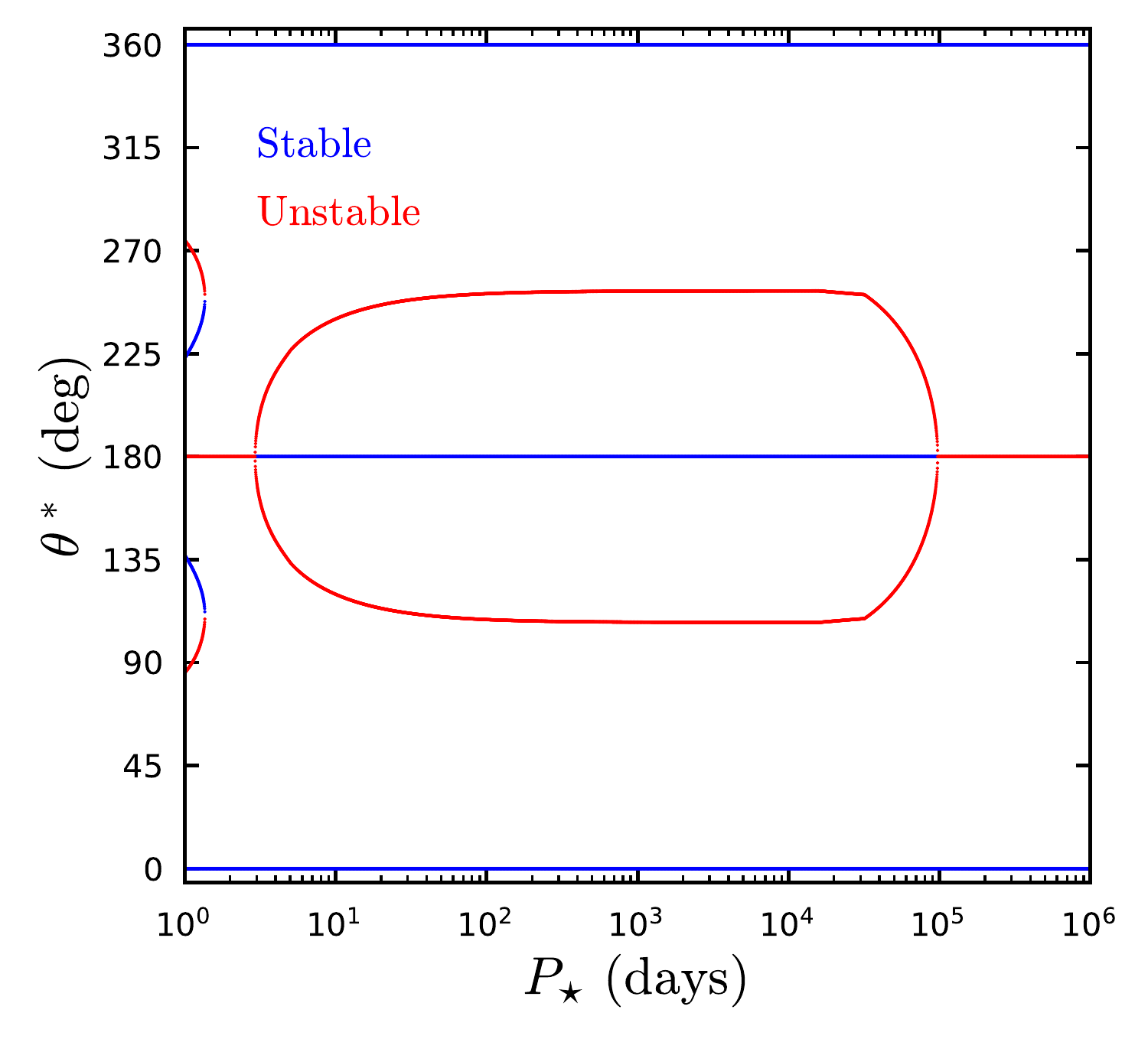}
\caption{Fixed points for the obliquity $\theta^*$ (see also Fig.~\ref{fig:theta_dot_vs_theta}) as a function of the stellar spin period. We have set the planet orbital period to $P_{\rm orb} = 2$ days, and the tidal parameters to $Q_0' = 10^6$ and $p = 3$. We include all values of $(m,m')$ that contribute to the torque. For $P_\star \gtrsim 2.9$ days, an additional unstable fixed point emerges, causing the anti-aligned state to become stable for a wide range of stellar spin periods. The additional fixed points that emerge for $P_\star \lesssim 1.3$ days are largely irrelevant in typical hot Jupiter systems}
\label{fig:bifurcation}
\end{figure}

Figure \ref{fig:bifurcation} illustrates the behavior of $\theta^*$ as a function of $P_\star$. The number of $\theta^*$ and their stability depends upon the stellar spin period (with all other system parameters held fixed). The spin-orbit aligned state is always stable, while the anti-aligned state may be stable or unstable depending upon the spin period. This indicates that the obliquity may spend periods of time locked at $180^\circ$ (as we show later via numerical integrations). 

Even if the obliquity reaches one of the non-zero $\theta^*$ states, this does not necessarily imply that the system will remain in such a state for long, since $\theta^*$ and its stability depends on the spin and orbital period, which both continue to evolve when $\theta = \theta^*$. For sufficiently large obliquities, $\dot{\Omega}_\star < 0$, causing the star to spin down even as the orbital period shrinks. When $\theta = 180^\circ$, the stellar spin magnitude evolves according to
\be
\frac{\D \Omega_\star}{\D t} = -\frac{6 \pi T_0}{5 I_\star} \Delta'_{-22} - \frac{T_{\rm MB}}{I_\star} < 0.
\label{eq:dot_omegas_antiparallel}
\ee
Since $\Delta_{-22} > 0$, the star's rotation must slow down
in the  anti-aligned state.

\begin{figure*}
\centering 
\includegraphics[width=\textwidth]{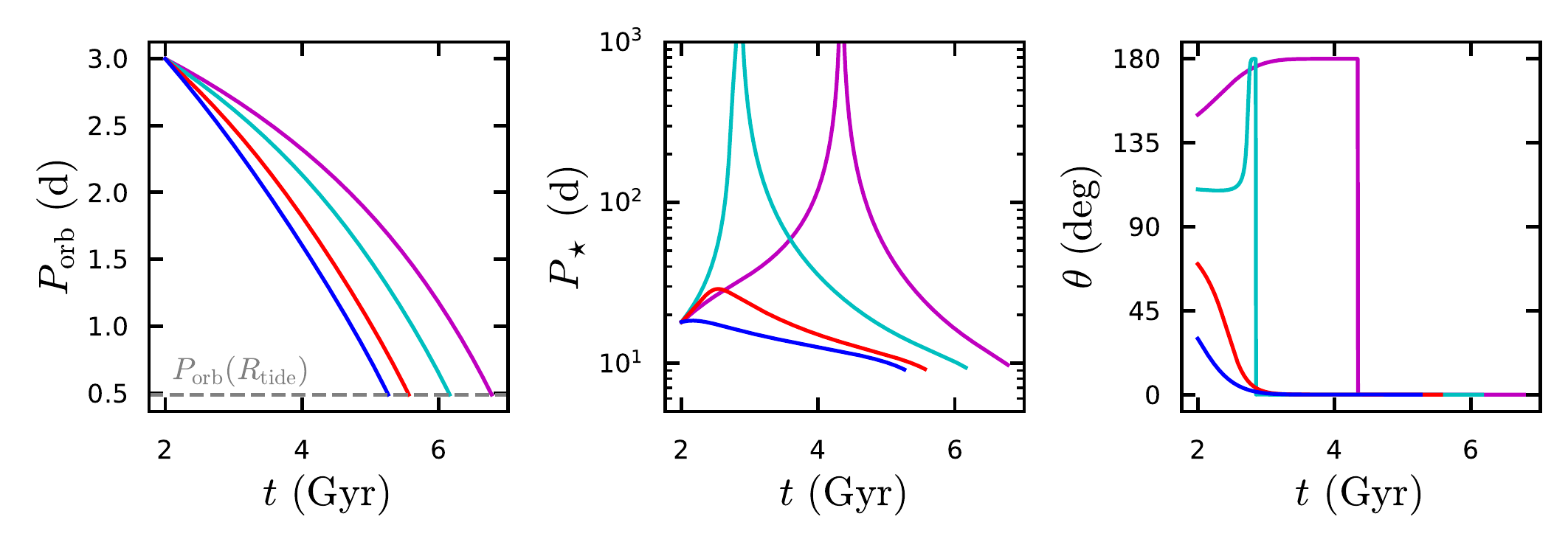}
\caption{Examples of spin-orbit evolution for a Jupiter mass and radius planet orbiting a solar mass and radius star, starting from various initial obliquities, and with the tidal parameters set to $\log Q_0' = 6.5$ and $p = 3$. The planet's arrival time and initial orbital period are 3 days and 2 Gyr respectively, appropriate for a hot Jupiter that formed due to high-eccentricity migration. All systems spend $\sim$2 Gyr or more in a spin-orbit aligned state, before eventually reaching the tidal disruption radius. Systems with initially prograde obliquities are driven towards $0^\circ$. Systems with initially retrograde obliquities are first driven towards $180^\circ$, as the stellar spin rate continuously decreases (see equation \ref{eq:dot_omegas_antiparallel}). Eventually the star halts, before spinning up in the opposite direction, causing $\theta$ to flip from $180^\circ$ to $0^\circ$. }
\label{fig:examples}
\end{figure*}

Figure \ref{fig:examples} depicts the spin and orbital evolution as a function of time for a range of different initial obliquities, spanning both initially prograde and retrograde orientations. The initial orbital period and arrival time for the planet are held fixed at 3 days and 2 Gyr respectively, which is consistent with a hot Jupiter that previously underwent high-eccentricity migration. We assume an initial stellar spin period of 3 days, and evolve the spin magnitude due to magnetic braking alone up to the time of the planet's arrival, and then evolve the system under the influence of both tides and magnetic braking. In these examples, the orbital evolution is similar in all cases, but the spin and obliquity evolution differs for the initially prograde and retrograde systems. For the initially prograde systems, the spin frequency briefly decreases due to magnetic braking, before the (positive) tidal torque overcomes the negative spin-down torque, thus spinning up the host star. Meanwhile, the obliquity is driven towards alignment. For the initially retrograde systems, the star is de-spun while the obliquity is temporarily driven towards anti-alignment. Then, the star halts momentarily, before spinning up in the opposite direction. As a result, the obliquity flips from $180^\circ$ to $0^\circ$. Systems with initially retrograde obliquities experience slower orbital decay than those with initially prograde obliquities. This in contrast with previous studies employing a constant time lag (in which $Q'$ decreases with forcing frequency) \citep[e.g.][]{barker2009}. This behavior in our
model is a consequence of the rise in $Q'$ with forcing frequency.

In all of the examples depicted in Fig.~\ref{fig:examples}, the star spends the majority
of the time in a spin-orbit aligned state. The same initial conditions, evolved under a constant phase-lag formalism (rather than frequency-dependent), fail to realign before the planet reaches the tidal disruption radius.

\section{Parameter Exploration}
\label{sec:parameter}
Having demonstrated that a frequency-dependent tidal quality factor can successfully re-align obliquities before destroying planets (see Fig.~\ref{fig:examples}), we next explore the spin and orbital evolution for a large number of initial conditions, encompassing different possible dynamical histories for the planet. In addition to varying the initial conditions for the planetary orbit, we also vary the free parameters in the tidal model, including the exponent $p$, and the scaling of the quality factor $Q'_0$.

We generate a population of 5000 planets with mass $1 \mjup$ and radius $1 \rjup$ orbiting a solar-type host star. We sample the initial spin period of the star (during the T-Tauri phase) in the range from 1 to 10 days and the orbital period in the range from 0.5 to 6 days, where the lower bound is chosen to avoid tidal disruption. We conduct separate experiments assuming different arrival times for the planet: an early arrival at $1$ Myr (denoted as \texttt{EARLY}), and a late arrival at $1$ Gyr (denoted as \texttt{LATE}). For the late-arriving planets, the stellar spin has de-spun to a period of 12 to 16 days at the time of the planet's arrival. Although the obliquity distributions produced by models of planet migration are often non-isotropic \citep[e.g.][]{anderson2016,vick2019,teyssandier2019}, we assume an initially isotropic distribution for simplicity. Furthermore, our goal here is to evaluate the prospects for obliquity realignment starting from a wide range of initial misalignments, rather than evolving the obliquities produced by specific models of hot Jupiter migration.

In each set of numerical experiments, we first evolve the system under magnetic braking up to the time of the planet's arrival. Once the planet is parked on a short-period orbit, we then evolve the system under both tides and magnetic braking up to a randomly chosen time in the range $1 - 10$ Gyr.

\begin{figure*}
\centering 
\includegraphics[width=0.9\textwidth]{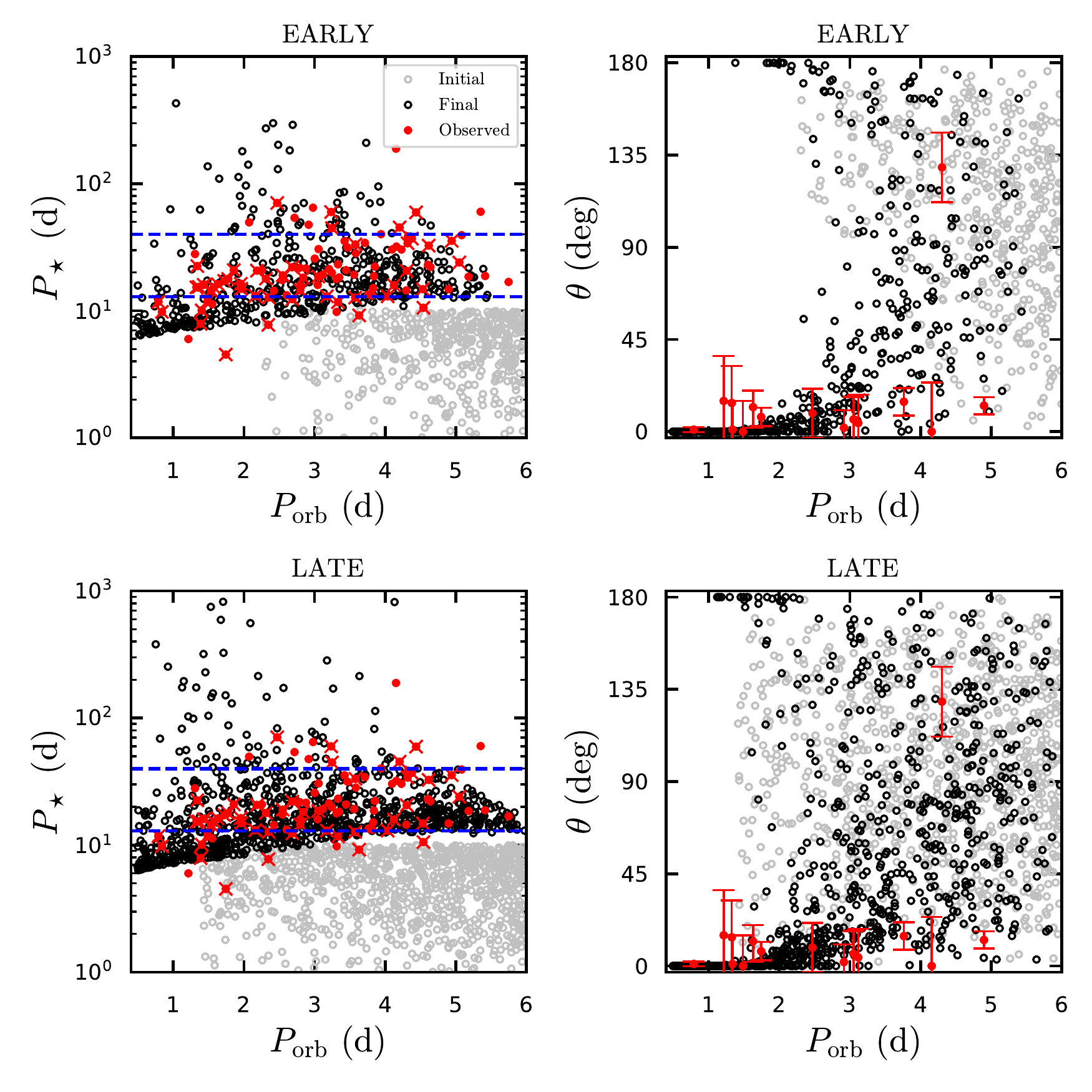}
\caption{Results of a large number of numerical integrations of star-planet systems, consisting of a Jupiter-mass and radius planet orbiting a solar-type star. We conduct different sets of numerical experiments, allowing the planet to arrive at a time $1$ Myr (denoted as \texttt{EARLY}), and a time $1$ Gyr (denoted as \texttt{LATE}). For each set of numerical experiments, we sample the initial orbital period in the range ($0.5 - 6$) days, the initial spin-period of the star (during the T-Tauri phase) in the range $(1 - 10)$ days, and the initial obliquity from an isotropic distribution. We integrate the system up to a time randomly chosen in the range ($1 - 10$) Gyr. The red points indicate observed systems; crosses indicate systems for which the spin period was measured photometrically, and circles indicate systems where the spin period was
calculated from estimates of $v \sin i$ and stellar radius. The black and gray circles indicate the simulation results for the surviving planets (satisfying $a > \rtide$). Gray circles indicate the initial values, while black circles indicate the values at a randomly selected time. The blue dashed lines indicate the expected range of spin periods for isolated stars (lacking tidal interactions), and are included for reference.}
\label{fig:pstar_porb}
\end{figure*}

We carry out both the \texttt{EARLY} and \texttt{LATE} experiments, fixing the tidal parameters to the fiducial values $p = 3$ and $Q_0' = 10^6$. The results are depicted in Fig.~\ref{fig:pstar_porb}, along with observed systems. The selection of these observed systems is described in detail in Section \ref{sec:observed}, consisting of transiting planets with masses greater than $0.5 \mjup$ orbiting stars with effective temperatures in the range from 5500 to 6000\,K. We supplement these transiting systems with additional non-transiting planets in the same mass and temperature range, taken from an upcoming study of stellar rotation in hot Jupiter systems (Tejada et al. 2021, in preparation). The rotation period is the photometric period when available, and otherwise obtained from the $v \sin i$ and stellar radius assuming $\sin i=1$. The simulated systems roughly match the observed systems in the $P_{\star}$-$P_{\rm orb}$ plane, although a closer match could possibly be obtained with different initial conditions and planet properties. In particular, the most rapidly rotating star in our sample is CoRoT-2, with a photometric rotation period of only 4.5 days -- our simulations do not quite reproduce this system in the $P_\star$-$P_{\rm orb}$ plane. Since the mass of CoRoT-2b is $\sim 3.5 \mjup$, and we have assumed a planet mass of only $1 \mjup$, better agreement for this system would be obtained by increasing the planet mass in our simulations. The dashed lines in the left panels of Fig.~\ref{fig:pstar_porb} bracket the range of spin periods expected due to magnetic braking alone. Systems outside of these boundaries have therefore undergone tidal evolution, leading to either spin-up or spin-down, depending on the initial conditions, and especially on the initial obliquity. 

The right panels of Fig.~\ref{fig:pstar_porb} depict the obliquity versus the planetary orbital period. For the \texttt{EARLY} planets, systems with orbital periods less than about $2$ days show a strong preference for alignment, with a small fraction of anti-aligned systems. The \texttt{LATE} planets exhibit a similar trend within an orbital period of about $1.75$ days.

\begin{figure*}
\centering 
\includegraphics[width=\textwidth]{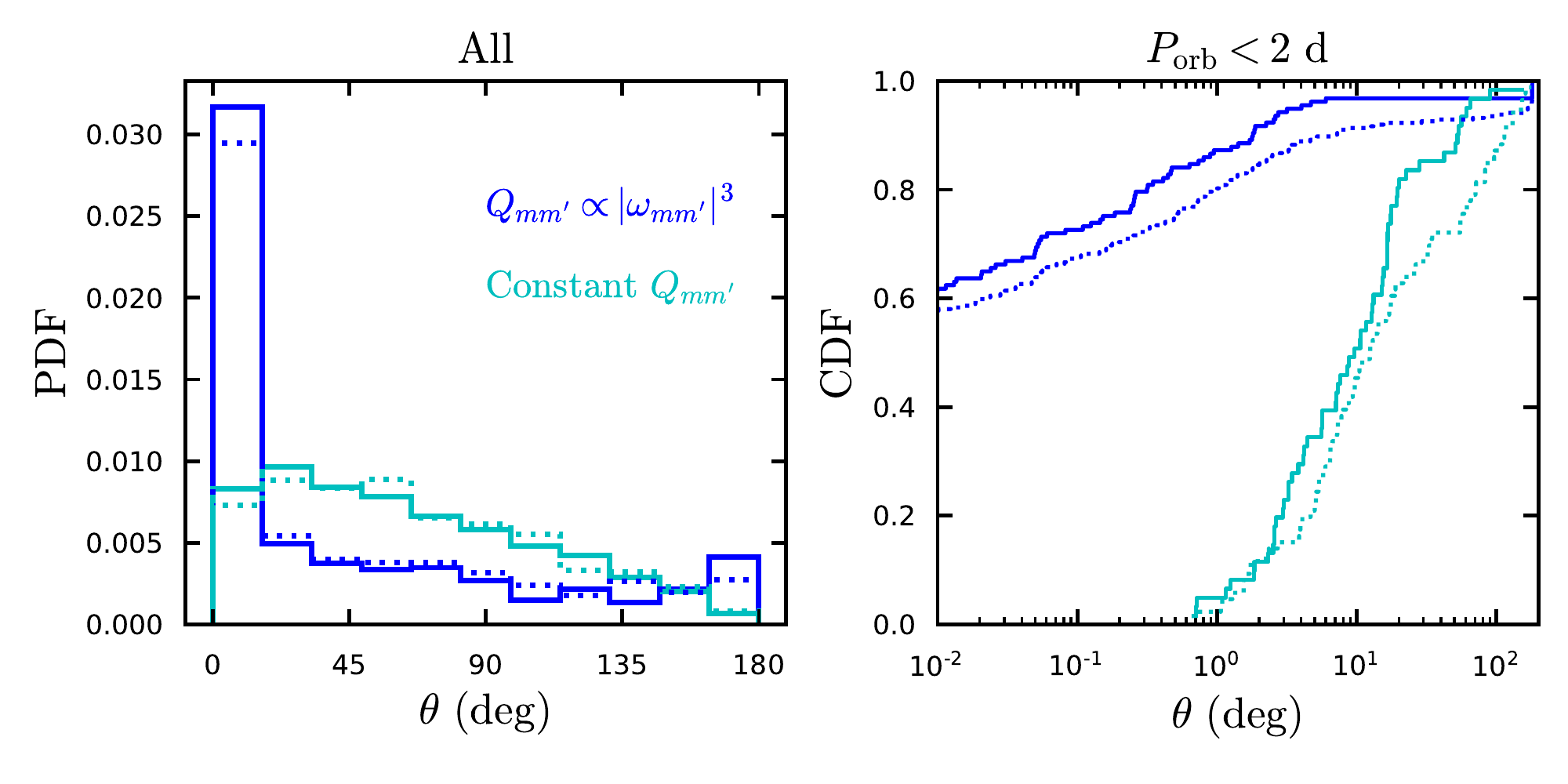}
\caption{Distribution of obliquities following tidal evolution. The left panel shows the obliquity distribution for all surviving planets (satisfying $a > \rtide$), and the right panel shows the cumulative distribution for the surviving planets with orbital periods less than 2 days. The blue lines indicate our fiducial tidal model, and the cyan lines depict a tidal model with constant phase lag for reference. The solid lines indicate the \texttt{EARLY} simulations, and the dotted lines indicate the \texttt{LATE} simulations.}
\label{fig:theta_comp_p_tarr}
\end{figure*}

Next, we repeat the \texttt{EARLY} and \texttt{LATE} experiments assuming a constant quality factor ($p = 0$ in our tidal model, and choosing $Q_0' = 10^6$). Figure \ref{fig:theta_comp_p_tarr} compares the predictions of the frequency-dependent quality factor ($p = 3$) and constant quality factor ($p = 0$). The left panel depicts the obliquity distributions for all the surviving planets, and the right panel shows the cumulative distribution for all surviving planets within $2$ days. The frequency dependent quality factor causes the planets within $2$ days to efficiently realign, with $\sim 80 \%$ of these systems having obliquities less than $1^\circ$.

\begin{figure*}
\centering 
\includegraphics[width=0.9\textwidth]{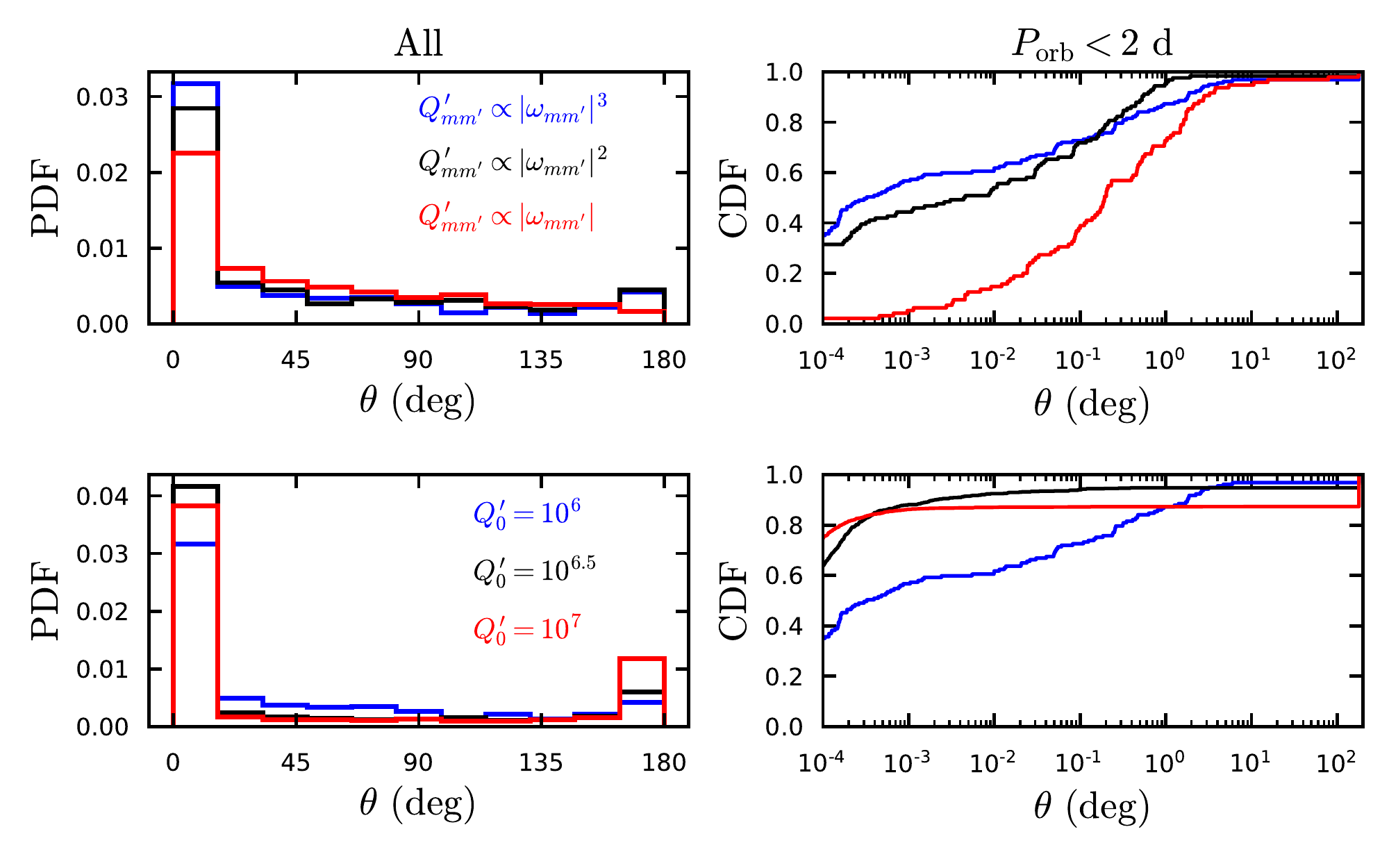}
\caption{Similar to Fig.~\ref{fig:theta_comp_p_tarr}, illustrating the dependence on the exponent $p$ and $Q_0'$. We show results only for the \texttt{EARLY} simulations. The top panels illustrates the implications of varying $p$ (with fixed $Q_0' = 10^6$), and the bottom panels illustrate varying $Q_0'$ (with fixed $p = 3$). Even a quality factor that scales only linearly with forcing frequency ($p = 1$) causes efficient obliquity realignment, indicating that the results are not overly sensitive to the exact scaling.}
\label{fig:theta_comp_p_Q}
\end{figure*}

Finally, in Fig.~\ref{fig:theta_comp_p_Q} we demonstrate in detail how the results depend on the tidal parameters $p$ and $Q_0'$. We show results only for the \texttt{EARLY} simulations, as similar results were obtained for the \texttt{LATE} simulations. We find that
the ability to realign systems before tidal destruction
is not overly sensitive to these tidal parameters.
Even if the quality factor depends only linearly on tidal frequency ($p = 1$) realignment is efficient, with $\sim$70\% of systems within $2$ days exhibiting obliquity alignment to within less than $1^\circ$. Increasing $Q_0'$ increases the efficiency of obliquity realignment.

\section{Comparison with Observed Hot Jupiters}
\label{sec:observed}

\subsection{Sample Selection and Stellar Age Determination}
In this section we calculate possible tidal histories for real hot Jupiters orbiting cool host stars. Since the rotational evolution of the star depends on the extent of the surface convective region, and because magnetic braking is best understood for solar-like stars, we restrict our attention to systems with temperatures in the range $(5500-6000)$K. In order to construct as homogeneous a sample as possible, we adopt the effective temperatures and metallicities from SWEET-Cat \citep{santos2013}, which is regularly updated with new planet discoveries. We then cross-match with planets from the NASA Exoplanet Archive and select transiting planets with masses larger than $0.5 \mjup$ and orbital periods less than 5 days. The reason we focus on transiting planets is that their true masses are known,
as opposed to Doppler planets for which only the minimum mass is known.  The upper limit on the orbital period was chosen to ensure that the sample largely contains systems susceptible to tidal evolution.  We reject any systems with an eccentricity above $0.05$, to avoid introducing additional parameters in the model to account
for tides raised on the planet by the star. We note that many of the planets in our sample have orbital eccentricities that are poorly
constrained by the available data. However, observations have
shown that highly eccentric systems around cool stars are rare when the period is shorter than 5 days. For each system, we perform a literature search to determine the stellar rotation period, obtained either from photometric variability (when available), or from the stellar $v \sin i$ and radius (assuming spin-orbit alignment). We discard systems lacking any information on the spin period.

We use the \texttt{Isochrones} package to calculate stellar ages \citep{morton2015}. \texttt{Isochrones} provides Bayesian stellar age and mass estimates by interpolating over stellar isochrones obtained from the MIST models \citep{choi2016, dotter2016}. \texttt{Isochrones} accepts a variety of user inputs, including any combination of effective temperature, metallicity, surface gravity, stellar density, V-band extinction, parallax, and photometry. We input the stellar effective temperature and metallicity from SWEET-Cat, along with the stellar density obtained from the transit light curve parameters $P_{\rm orb}$ and $a/R_\star$, via
\be
\rho_\star = \frac{3 \pi}{G P_{\rm orb}^2} \bigg(\frac{a}{R_\star} \bigg)^3.
\label{eq:density}
\ee
We also input the Gaia parallax and photometry ($G, BP, RP$), and the 2MASS ${\rm JHK_s}$ and WISE 123 bands. We use the 3D galactic extinction model MWDUST \citep{bovy2016} to estimate the extinction at the distance implied by the Gaia parallax, and impose a flat prior with maximum $A_{\rm V,max} + 0.1 \ {\rm mag}$, where $A_{\rm V,max}$ is the maximum extinction along the entire line of sight. We allow for additional extinction of $0.1$ mag beyond the maximum value to account for uncertainties in the galactic extinction model. See Appendix \ref{sec:ages} for additional details.

After discarding several systems with unreliable parameters or output ages (see Appendix \ref{sec:ages}), our final sample consists of 46 hot Jupiters orbiting cool host stars. This sample is a subset of the observed planets depicted in Fig.~\ref{fig:pstar_porb}, since here we restrict our attention to transiting planets.

\subsection{Tidal Histories for Hot Jupiters Orbiting Cool Stars}
\label{sec:histories}
\begin{figure}
\centering 
\includegraphics[scale=0.47]{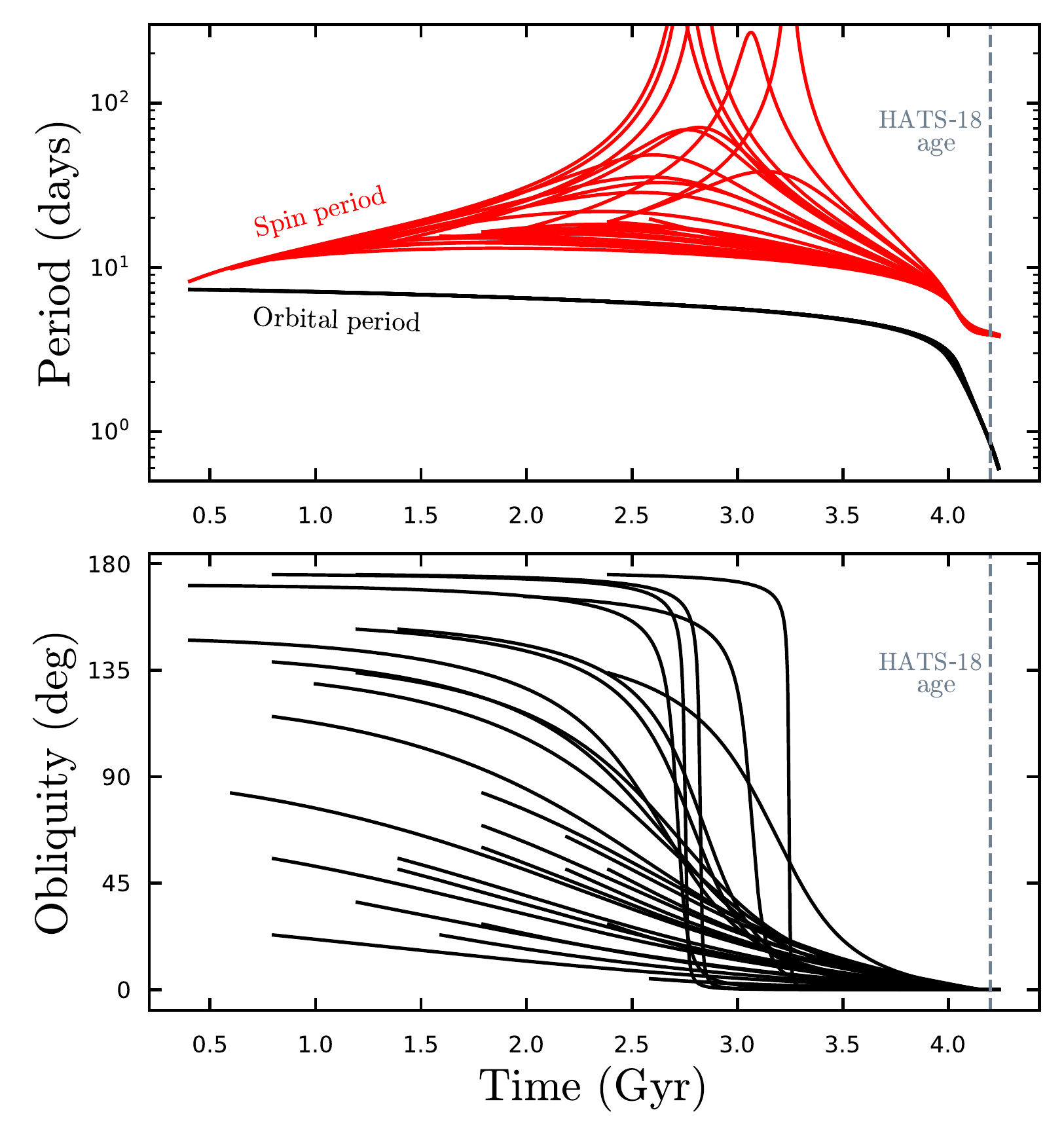}
\caption{Possible tidal histories for the HATS-18 system, showing a variety of initial obliquities and planet arrival times, and adopting a system age of $4.2$ Gyr. Regardless of the initial obliquity and arrival time, the obliquity has decayed close to $0^\circ$ at the observed system age.}
\label{fig:hats18}
\end{figure}

Next, we present tidal histories for the sample of hot Jupiters orbiting cool stars, adopting the fiducial scaling of the quality factor with forcing frequency ($p = 3$). Figure \ref{fig:hats18} depicts possible tidal histories for the HATS-18 system \citep{penev2016}, an illustrative case.
We assumed the age to be $4.2$ Gyr, and sampled a range of initial obliquities and arrival times for the planet. We then iteratively determined the values of the initial orbital period and $Q_0'$ that reproduce the observed spin and orbital periods at the adopted system age. For all of the initial conditions explored, the obliquity has realigned at the adopted system age.

\begin{figure*}
\centering 
\includegraphics[width=\textwidth]{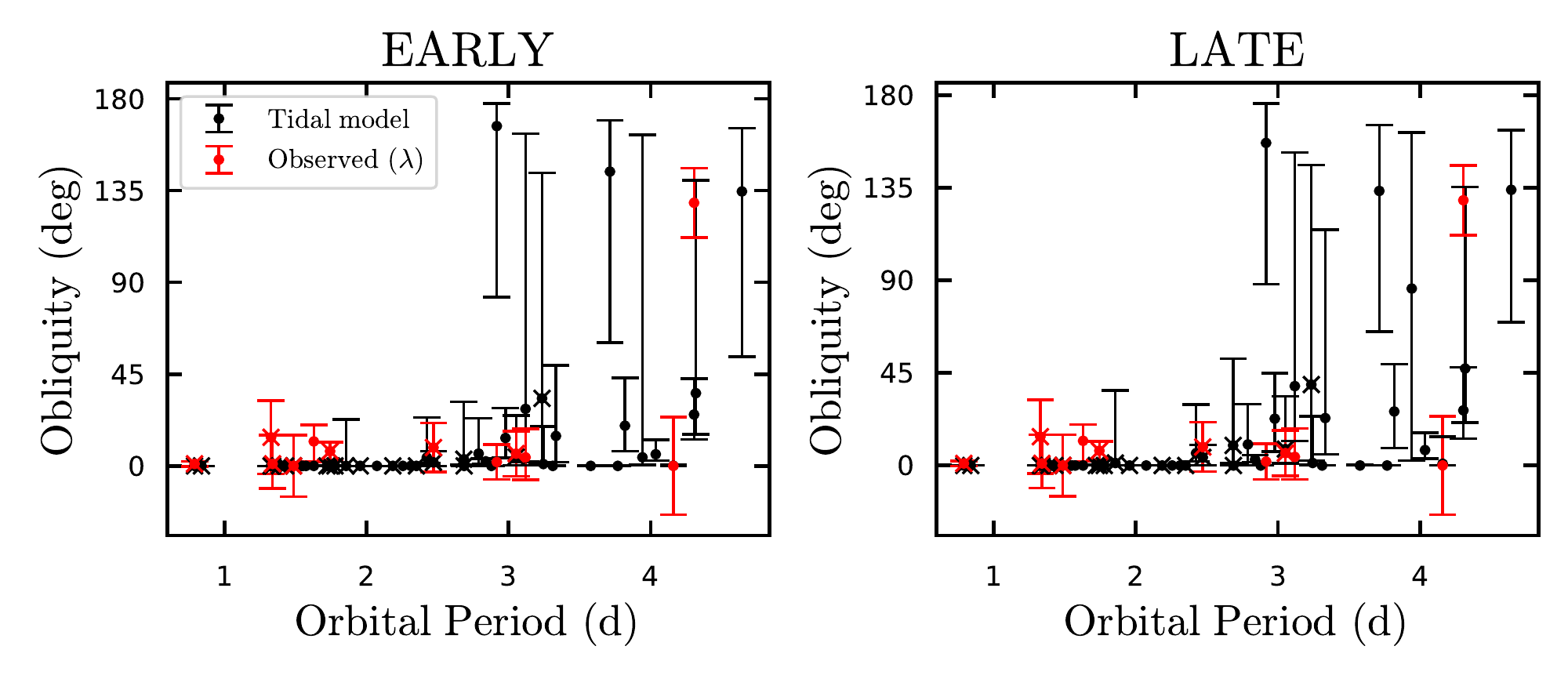}
\caption{Results of tidal evolution calculations for the entire sample of 46 hot Jupiters orbiting cool stars. We plot the final obliquity distributions versus the orbital period, based on
the set of initial conditions that led to agreement with
the observed spin and orbital periods (within their uncertainties). 
The black error bars indicate the median and 16th and 84th percentiles of the obliquity posteriors from our simulations.
The red points indicate observed sky-projected obliquities. Points with an 'x' indicate the systems for which the spin period
has been measured from photometric variations. Our tidal model predicts that hot Jupiters with orbital periods less than $\sim 2.5$ days should have obliquities less than $1^\circ$.}
\label{fig:observed}
\end{figure*}

Similar to the example depicted in Fig.~\ref{fig:hats18} for HATS-18, we calculate possible tidal histories for each of the 46 observed hot Jupiters, sampling the stellar age directly from the \texttt{Isochrones} age posteriors. We conduct two separate experiments, corresponding to early arrival of the planet due to disk migration, and late arrival due to high-eccentricity migration. We refer to the two experiments as \texttt{EARLY} and \texttt{LATE} respectively. The procedure is as follows:
\begin{itemize}
\item Draw a random sample $t_{\rm age}$ from the \texttt{Isochrones} age posterior.

\item Sample an arrival time $t_{\rm arr}$ for the planet (3 Myr for the \texttt{EARLY} experiment, and uniformly in the range between
$50 {\rm Myr}$ and $0.9 t_{\rm age}$ for the \texttt{LATE} experiment).

\item Sample the initial obliquity from an isotropic distribution (uniform in $\cos \theta$).

\item Sample an initial stellar T-Tauri spin period uniformly between (1-10) days, and evolve the spin period due to magnetic braking up to the time of the planet's arrival.

\item Evolve the system due to both tides and magnetic braking up to $t_{\rm age}$, iteratively adjusting the initial orbital period and the value of $Q_0'$ to reproduce the observed spin and orbital periods at $t_{\rm age}$ (within the observational uncertainties). We impose a minimum uncertainty of $10 \%$ for the spin period,
regardless of the quoted observational uncertainty.

\end{itemize}
For each system, this procedure results in a distribution of stellar obliquities at the current system age. Fig.~\ref{fig:observed} illustrates the final obliquity distributions from our simulations of each of the 46 systems, as a function of the observed orbital period. The black error bars indicate the 16th and 84th percentiles of the obliquity distribution, while red error bars indicate the observed sky-projected obliquities for the systems for which obliquity measurements are available. Our calculations generally imply that systems with orbital periods less than $2.5$ days have undergone substantial obliquity damping, and have final obliquities smaller than
$1^\circ$.

\begin{figure}
\centering 
\includegraphics[scale=0.47]{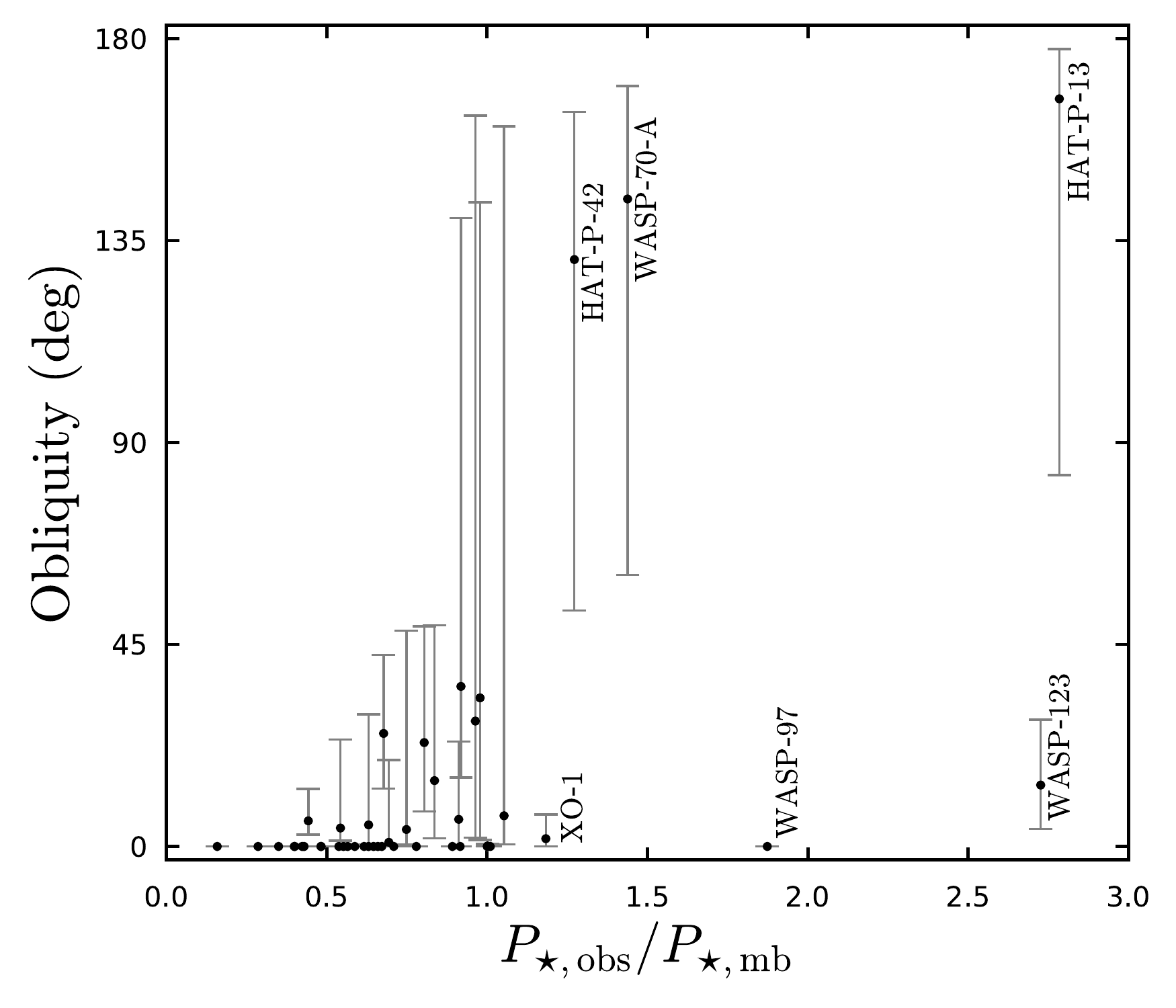}
\caption{Obliquity distribution as a function of $P_{\star, \rm obs} / P_{\star, \rm mb}$, where $P_{\star, \rm obs}$ is the observed spin period and $P_{\star, \rm mb}$ is the expected spin period of an isolated star at the system age (see equation \ref{eq:Pmb}, and note that we use the median value over our initial conditions and age posteriors). We show results only for the \texttt{EARLY} simulations (similar results are obtained for the \texttt{LATE} simulations), and include only the initial conditions that reproduced the observed spin and orbital periods (within our adopted uncertainties). The systems showing a preference for retrograde obliquities have $P_{\star, \rm obs} / P_{\star, \rm mb} \gtrsim 1$. This feature arises because initially retrograde configurations act to de-spin the host star.}.
\label{fig:observed_pspin}
\end{figure}

For the majority of systems, the current obliquity is predicted
to be close to zero, indicating efficient realignment. However, for several stars that are observed to be slowly rotating, our calculations predict retrograde obliquities. To demonstrate this, we compare the observed spin period of the star to the expected spin period of an isolated star at the same age (for which the spin period evolves due to magnetic braking alone). The spin period as a function of age for an isolated star is approximately
\be
P_{\star,\rm mb} \simeq P_{\star,0} \bigg(1 + \frac{8 \pi^2 \alpha t_{\rm age}}{P_{\star,0}^2} \bigg)^{1/2},
\label{eq:Pmb}
\ee
where $P_{\star,0}$ is the spin period during the T-Tauri phase, and $\alpha$ is given by equation (\ref{eq:alpha}). Thus, for each system we can compare $P_{\star,\rm obs}$ and $P_{\star,\rm mb}$. Fig.~\ref{fig:observed_pspin} shows the final obliquity distribution versus $P_{\star, \rm obs} / P_{\star, \rm mb}$, where we have used the median value of $P_{\star,\rm mb}$ obtained from our initial spin periods and age posteriors. For the slowly rotating stars ($P_{\star, \rm obs} / P_{\star, \rm mb} \gtrsim 1$), there is often a preference for initially retrograde obliquities. This preference may be understood by noticing that tides in sufficiently misaligned configurations act to de-spin the star (see, e.g. equation \ref{eq:dot_omegas_antiparallel}). However, we stop short
of predicting that such stars are truly retrograde, because
all of the rotation periods for the anomalously slowly rotating stars were derived from the stellar $v \sin i$ under the
assumption $\sin i=1$, rather than from the less ambiguous method of detecting a periodicity in the photometric variability. As a result, another possible interpretation is that the spin period has been overestimated because $\sin i \neq 1$, i.e., the obliquity
is currently large.  In this scenario, repeating the tidal calculations with the (unknown) true spin period would
possibly allow initially prograde configurations to reproduce the system properties, by decreasing or eliminating the need for tidal de-spinning. Given our present knowledge, we cannot distinguish between these two possibilities.

\subsection{Remaining Lifetimes}
\label{sec:lifetimes}
\begin{figure}
\centering 
\includegraphics[scale=0.475]{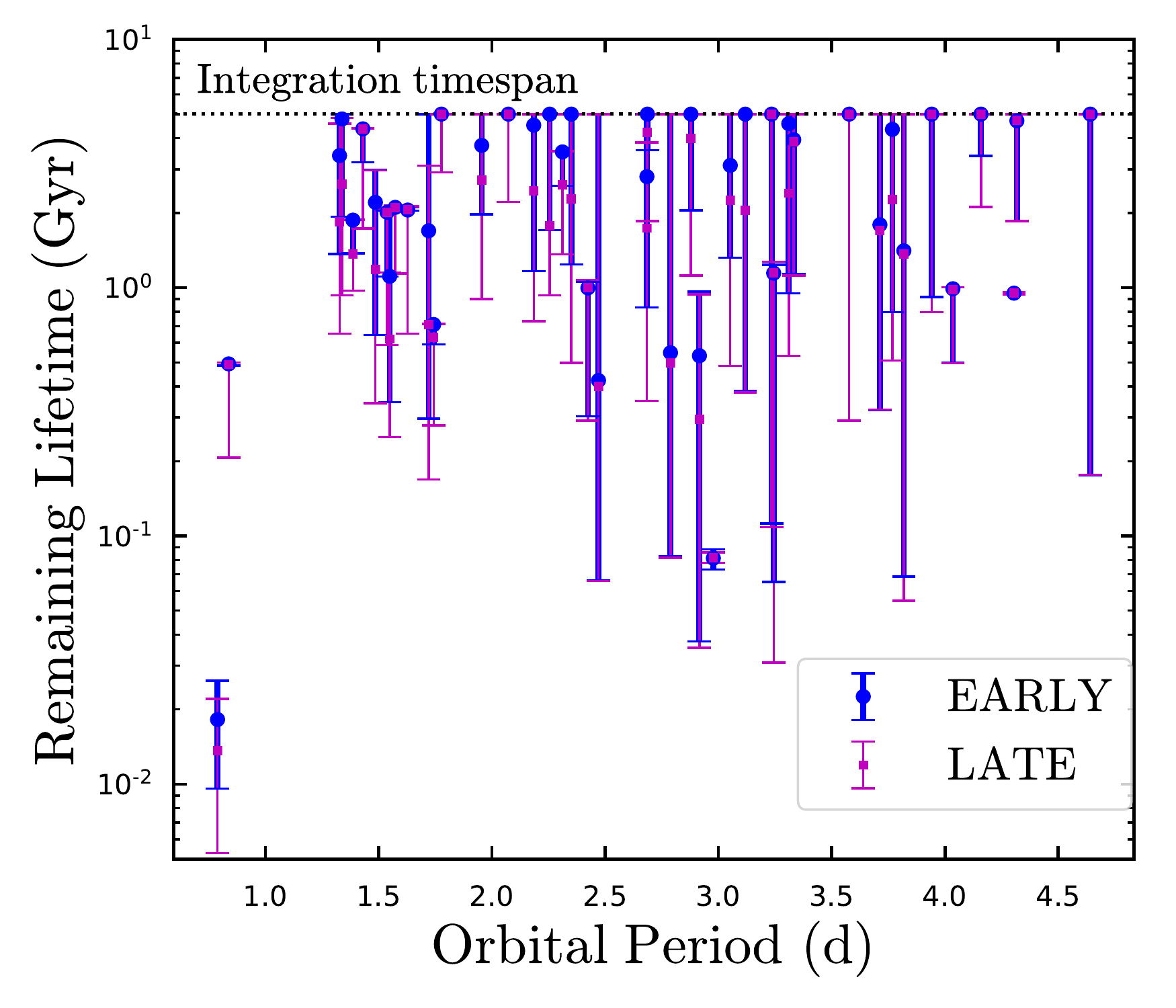}
\caption{Remaining lifetime for the sample of 46 hot Jupiters orbiting cool host stars. For each system, we evolve the spin and orbit from the observed state until the planet crosses $\rtide$ (with $\eta = 2.5$; see equation \ref{eq:rtide}), or until $5$ Gyrs have elapsed. If the planet remains exterior to $\rtide$ after 5 Gyr, we set the remaining lifetime to be equal to 5 Gyr to obtain a lower limit. The calculations reveal that shortest period planet (WASP-19b) will cross $\rtide$ within $\sim 10$ Myr, due to its current proximity to $\rtide$. All other planets will survive for at least hundreds of Myr.}
\label{fig:lifetime}
\end{figure}

We also calculate posteriors for the remaining lifetime, defined as the time until the planet crosses the tidal disruption radius $\rtide$, given by equation \ref{eq:rtide} (with $\eta = 2.5$). Each set of initial conditions (with $t_{\rm arr}$, $\theta_0$, and $t_{\rm age}$ randomly sampled) is integrated forward in time up to 5 Gyr, using the values of $Q_0'$ and the initial
orbital period that were previously determined to reproduce the observed spin and orbital periods. If the planet crosses $\rtide$ we terminate the integration and record the remaining lifetime. If the planet remains exterior to $\rtide$ over the entire 5 Gyr integration, we set the remaining lifetime to $5$ Gyr. Fig.~\ref{fig:lifetime} depicts the results as a function of orbital period. Most of the planets are not in immediate danger of being devoured by their host stars. The typical remaining lifetime is at least hundreds of millions years. An exception to this trend is WASP-19b (the shortest-period planet in our sample), which our calculations predict will cross $\rtide$ within 10--20\,Myr. In contrast, the only other planet in our sample with a similar orbital period (HATS-18b) has a remaining lifetime of $\sim 500$ Myr. The key difference between these two planets is that WASP-19b is located extremely close to $\rtide$, according
to our definition. This illustrates that the planet's fate depends sensitively on the chosen disruption criterion, through the value of $\eta$ (see equation \ref{eq:rtide}).

\begin{figure*}
\centering 
\includegraphics[width=\textwidth]{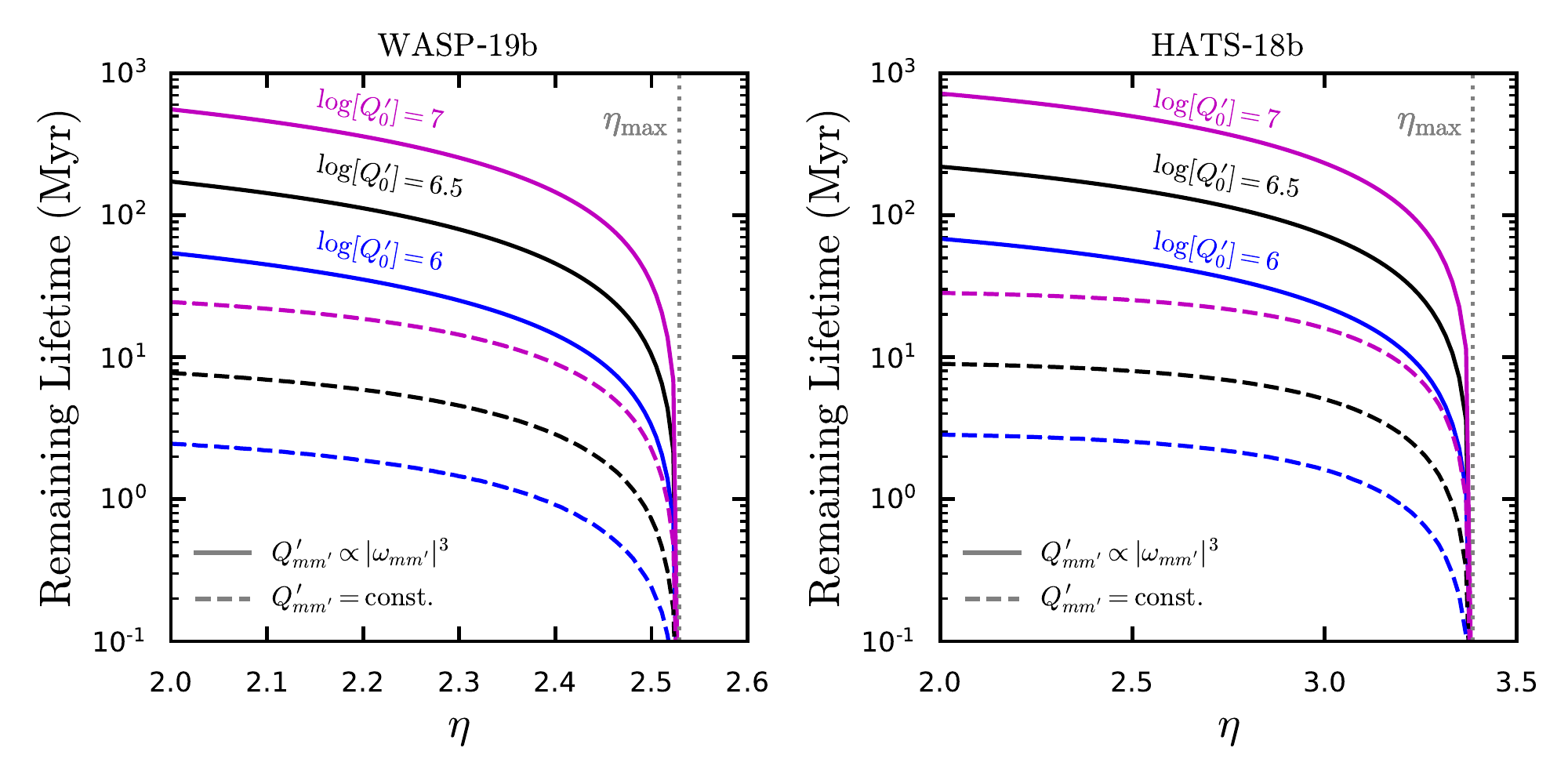}
\caption{Remaining lifetime as a function of the definition of the tidal disruption radius [with $\rtide = \eta R_{\rm p} (M_\star / M_{\rm p})^{1/3}$] for the two shortest-period planets in our sample. The left and right panels show results for WASP-19b and HATS-18b respectively. We vary $\eta$ between 2 and the maximum possible value by requiring that the current planet semi-major axis is larger than $\rtide$ (see equation \ref{eq:maxeta}). The blue, black, and magenta curves show $\log_{10}(Q_0') = 6,6.5,7$ and the solid and dashed curves show $p = 3$ (our fiducial value) and $p = 0$ (corresponding to constant $Q_{mm'}'$). Comparing the solid and dashed curves at fixed $\eta$ and $Q_0'$, the frequency-dependent quality factor prolongs the remaining lifetime by more than a factor of 10.}
\label{fig:lifetime_eta}
\end{figure*}

To explore this issue further, we perform additional calculations for the WASP-19 and HATS-18 systems, varying $\eta$ between 2 and a maximum possible value
\be
\eta_{\rm max} = \frac{a}{R_{\rm p}} \bigg(\frac{M_{\rm p}}{M_\star} \bigg)^{1/3}.
\label{eq:maxeta}
\ee
For each value of $\eta$, we integrate the system forward in time and calculate the remaining lifetime, assuming spin-orbit alignment and varying $Q_0'$. The results are shown in Fig.~\ref{fig:lifetime_eta}. For a given value of $Q_0'$, reducing $\eta$ by a small amount prolongs the remaining lifetime for both systems.  We also compare the results for our fiducial frequency-dependent quality factor (with $p = 3$, solid curves) to a constant quality factor ($p = 0$, dashed curves). Fig.~\ref{fig:lifetime} reveals that the frequency-dependent quality factor prolongs the remaining lifetime by more than a factor of 10.

\section{Conclusion}
\label{sec:conclusion}
\subsection{Summary of Results}
Tidal re-alignment remains a promising explanation for at least some aspects of the well-known pattern relating measurements of stellar obliquity
and the effective temperature of the host star \citep{winn2010}. This paper describes our investigations of the tidal evolution of spin-orbit misaligned systems with a frequency-dependent stellar tidal quality factor. Our model is motivated by the recent study of stellar rotation in hot Jupiter systems by \cite{penev2018}, who found evidence that the tidal quality factor increases strongly with forcing frequency (assuming spin-orbit alignment). For spin-orbit misaligned systems, there are multiple independent tidal Fourier components, each operating at a different forcing frequency (a linear combination of the orbital and spin frequencies) and each associated with a potentially different quality factor \citep{lai2012,ogilvie2014}. We have implemented a frequency-dependant quality factor for each tidal component. The tidal component operating at the stellar spin frequency affects the obliquity without introducing orbital decay. Since the strong frequency dependence causes the components with the lowest forcing frequencies to have the smallest quality factors, this model efficiently re-aligns obliquities before the planetary orbit catastrophically decays. 

Including the effects of both tides and magnetic braking, we have explored this model for simulated hot Jupiters with a range of properties, as well as for a sample of real
hot Jupiters orbiting cool host stars. Using Gaia DR2 data, we have derived stellar ages for the hot Jupiter host stars, and explored possible tidal histories, starting with a broad range of initial stellar obliquities and two very different possibilities for
the arrival times of the planet. Our calculations confirm that large obliquities are efficiently damped. The model predicts extremely low obliquities (typically much less than $1^\circ$) for hot Jupiters orbiting cool stars within 2.4 days (Fig.~\ref{fig:observed}).

\subsection{Discussion}

The most important assumption in our model is that the tidal quality factor rises with forcing frequency over the range of frequencies typical in hot Jupiter systems. This assumption was inspired by the earlier findings of \cite{penev2018}, who
investigated the possible tidal evolution histories of a large collection of
hot Jupiter systems.  A nearly contemporaneous population-level study of tidal dissipation in hot Jupiter
systems was performed by \cite{collier_cameron2018}, who did not report
evidence for a frequency-dependent tidal quality factor. This makes
it important to compare the two studies to see if there is any contradiction.

The Bayesian hierarchical analysis of \cite{collier_cameron2018} was based on the observed distribution of planetary orbital periods, and assumed an initially log-normal semi-major axis distribution. Those authors chose to analyze all of the planets in an online catalog of transiting planets called TEP-Cat.\footnote{https://www.astro.keele.ac.uk/jkt/tepcat/} They  divided the sample into two groups: systems with $0.5 < P_{\rm orb}/P_{\star} < 2$ (for which inertial waves can be excited), and systems with relative spin and orbital periods outside of the aforementioned range (for which they assumed the dissipation of the equilibrium tide is the dominant mechanism). They inferred a quality factor $Q' \sim 10^7$ for the inertial-wave systems and $Q' \sim 10^8$ for the equilibrium-tide systems.
In contrast, \cite{penev2018} restricted their sample to have orbital
periods shorter than 3 days, and host stars cooler than 6200\,K, narrower
ranges than in the sample of \cite{collier_cameron2018}. As a result of the short orbital period cutoff, none of \cite{penev2018} systems are unambiguously in the inertial-wave regime. The inertial-wave sample of \cite{collier_cameron2018} consists of planets with longer tidal forcing periods compared to the equilibrium-tide sample (so that the condition $0.5 < P_{\rm orb}/P_{\star} < 2$ is satisfied). As a result,
the findings of \cite{collier_cameron2018} are broadly consistent with the
finding of \cite{penev2018} that longer-period systems have lower values
of $Q'_\star$.

The results of this paper are subject to some uncertainties and caveats. Notably, we adopted the empirically-motivated scaling for the stellar quality factor for each of tidal component, as found by \cite{penev2018} for spin-orbit aligned systems. For systems with spin-orbit misalignment, it is far from obvious whether the dissipation introduced by the various tidal components should be expected to scale in the same way. In the face of this
uncertainty, we adopted the same quality factor for each tidal component, representing the simplest possible assumption. The theoretically expected dependence of $Q’$ on tidal frequency remains highly uncertain and depends on the dissipation mechanism under consideration.

Tidal dissipation in solar-type stars may arise from the equilibrium tide or from
dynamical tides (the excitation and damping of various kinds of waves inside the star). In the case of equilibrium tides, dissipation is thought to arise from damping of turbulent viscosity in convection zones \citep{zahn1966}. In this theory, the scaling of the quality factor with tidal frequency depends on the eddy viscosity. If the eddy viscosity is independent of forcing frequency, the quality factor scales as $Q \propto \omega^{-1}$ (constant lag time).  However, for tidal frequencies larger than the convective turnover frequency $\omega_{\rm c}$, the eddy viscosity is thought to be reduced below this expectation.  The exact form of the reduction factor has been disputed for decades \citep{zahn1966,zahn1977,goldreich1977,goodman1997}. \cite{zahn1966} argued for a reduction factor of $\omega/\omega_c$, which implies that the quality factor should scale as $Q \propto \omega^0$ (i.e., constant Q). In contrast, \cite{goldreich1977} argued for a reduction factor of $(\omega / \omega_{\rm c})^2$, giving $Q \propto \omega$. Since then, numerical simulations have
been undertaken to determine the frequency dependence of the quality factor \citep{penev2009,ogilvie2012}, with a recent study finding support for the quadratic
reduction factor \citep{duguid2020}. This is qualitatively consistent with the models explored in this paper, especially the case of our $p = 1$ calculations. However, although the scaling with frequency is in general agreement with our models, the numerical simulations lead to quality factors that are larger in absolute terms,
which would lead to negligible obliquity and spin evolution. Various uncertainties remain in the models, however, and further studies are needed.

Regarding dynamical tides, a leading contender for the orbital evolution in hot Jupiter systems is the excitation and damping of internal gravity waves. For the solar-type stars considered in this paper, these waves are excited at the core-envelope boundary, and propagate inwards in the radiative zone. The degree of dissipation depends strongly on whether or not these waves break at the center \citep{goodman1998,barker2010}, which itself depends on the planetary and stellar masses, as well as the stellar age (see Figs.\ 8 and 9 of \citealt{barker2020}). If the waves are fully damped, this mechanism predicts the scaling $Q’ \propto \omega^{-8/3}$, which is very different from the form adopted in this paper.  Since the critical mass for wave breaking decreases strongly with age, the results of \cite{barker2020} predict that the quality factor should decrease with age, eventually leading to planetary engulfment. However, it is not clear what fraction of the observed hot Jupiters are currently in this wave-breaking regime, especially given the large uncertainties in stellar ages. Further studies and more precise stellar ages are needed to clarify this issue.

The excitation and damping of inertial waves is another contender for driving the orbital and obliquity evolution in hot Jupiter systems, particularly during the pre-main-sequence stage of stellar evolution. The theory of inertial waves predicts a strong and complicated frequency-dependence of the quality factor.  Averaged over frequency, the theory predicts $Q’ \propto \Omega_\star^{-2}$ \citep[e.g.][]{ogilvie2007}. \cite{mathis2015} has argued for greatly enhanced levels of dissipation due to inertial waves during the pre-main-sequence phase, based on a frequency-averaged two-layer model derived by \cite{ogilvie2013}. Assuming spin-orbit alignment, and coupling the \cite{ogilvie2013} model to stellar and orbital evolution, \cite{bolmont2016} investigated the tidal evolution of various star-planet systems, with subsequent calculations performed by \cite{heller2019} in concert with disk torques. These studies have recently been improved upon by \cite{barker2020}, accounting for the realistic structure (density profile) of the star. For spin-orbit
misaligned systems, the complexity of the calculation increases dramatically due to the multiple Fourier components at work. Accounting for the changes in dissipation due to stellar evolution for spin-orbit misaligned systems is therefore beyond the scope of this paper \citep[but see][for an analysis of the dissipation introduced by the $(m,m') = (1,0)$ component in isolation]{lin2017}. In any case, the previous works, while intriguing, are subject to their own uncertainties and caveats. If we had
taken into account the possible increase in dissipation during the pre-main sequence phase, the results of our simulations involving early-arriving planets would probably
be very different.  Even without inertial waves enhancing the dissipation, pre-main-sequence evolution could still be important simply because of the larger stellar radius. This issue therefore depends critically on the timing of hot
Jupiter emplacement.  If the majority of hot Jupiters arrive at their observed orbital locations early, due to disk migration or {\it in situ} formation, some of the results of this paper would need to be modified. If most hot Jupiters arrive after the star has settled onto the main
sequence, possibly due to some form of high-eccentricity migration, our results are valid as they are.

Since the tidal dissipation is expected to increase as the star leaves the main sequence (due to the increased stellar radius, as well as the excitation and damping of inertial waves and internal gravity waves), the remaining lifetimes derived in Section \ref{sec:lifetimes} should be interpreted as maximum lifetimes.  Our calculations predict that most hot Jupiters are not in immediate danger of being tidally destructed, with remaining lifetimes ranging from hundreds of Myr to several Gyr or more (see Fig.~\ref{fig:lifetime}). Since evidence for hot Jupiters to be tidally destroyed has been found \citep{hamer2019}, this may indicate tidal destruction is often initiated by stellar evolution. Our calculations suggest that WASP-19b (the shortest-period planet in our sample) may be on the verge of tidal disruption, although this conclusion depends strongly on the choice of the tidal disruption criterion (see Fig.~\ref{fig:lifetime_eta}). So far, the WASP-12 system remains the only hot Jupiter for which the period has been observed to be decreasing with time \citep{Maciejewski+2016, yee2020}. We note, however, that WASP-12 is a hot star ($T_{\rm eff} \approx 6300 {\rm K}$) and might be a subgiant
\citep{Weinberg+2017,BaileyGoodman2019}.
As a result, the tidal model adopted in this paper cannot be applied to this system; the frequency dependence of the quality factor inferred by \cite{penev2018} was
based on observations of cooler main-sequence stars that spin down with age.

In addition, the magnetic braking model adopted in this paper is relatively simplistic. For example, while we have assumed the star
is rotating uniformly, some degree of core-envelope decoupling may be necessary to reproduce the observed rotation periods in open clusters of different ages \citep[e.g.][]{irwin2007}. Allowing for the core-envelope decoupling would decrease the effective moment of inertia of the star participating in the tidal interaction, leading to more efficient obliquity realignment. Since our tidal model already predicts extremely efficient obliquity realignment, accounting for core-envelope decoupling would make our conclusion even stronger.

As a result, regardless of the possibilities for enhanced tidal dissipation due to stellar evolution, and modifications to the magnetic braking model, the main result of this paper is likely
to persist even after these physical effects are taken into account. Our tidal model implies that most hot Jupiters with orbital periods less than several days should have obliquities much smaller than $1^\circ$. These values are smaller than those predicted by a tidal model with a constant stellar quality factor (see the left-hand panel of Fig.~\ref{fig:theta_comp_p_tarr}), as well as
being smaller than the $6^\circ$ obliquity of the Sun. Our prediction may be testable through future high-precision Rossiter-McLaughlin measurements.

This paper has been motivated by the observed obliquity-temperature for hot Jupiters. While we have shown that our tidal model efficiently damps obliquities at short orbital periods ($< 3$ days), at longer periods the tidal damping is not efficient enough to erase large obliquities (see Fig.~\ref{fig:pstar_porb}). The typically low obliquities of planets orbiting cool stars even at longer orbital periods may instead reflect the formation conditions of these planets, or tidal effects not captured in our model.

We find it intriguing that some hot-Jupiter hosts
may have been tidally de-spun, rather than spun up, if the
initial obliquities were retrograde.  In such cases,
the obliquity may be temporarily driven towards $180^\circ$,
at which point the spin rate continues to decrease due to tides and magnetic braking. Eventually the star comes to rest, before spinning up in the opposite direction (see Fig.~\ref{fig:examples}). Thus, stars with anomalously slow rotation periods for their ages may be a clue that the obliquity was initially large.
There are a few cases in our sample that appear to be in
this category, subject to an ambiguity in the current spin
periods.  It would be interesting to measure their rotation periods more
directly, i.e., based on periodic photometric variations, and
test the notion that the planets have tidally de-spun their stars.

\acknowledgments
KRA thanks Joel Hartman for assistance with the stellar age calculations, and Roberto Tejada Arevalo for sharing his compilation of stellar rotation periods for use in Fig.~\ref{fig:pstar_porb}.  We also thank the referee, Adrian Barker, for constructive criticism that significantly improved the paper. This research has been supported in part by NASA grant ATP 80NSSC18K1009. KRA is supported by a Lyman Spitzer, Jr.~Postdoctoral Fellowship at Princeton University. The simulations presented in this paper were performed on computational resources managed and supported by Princeton Research Computing, a consortium of groups including the Princeton Institute for Computational Science and Engineering (PICSciE) and the Office of Information Technology's High Performance Computing Center and Visualization Laboratory at Princeton University.

\bibliography{refs}

\begin{thebibliography}{}
\expandafter\ifx\csname natexlab\endcsname\relax\def\natexlab#1{#1}\fi
\providecommand{\url}[1]{\href{#1}{#1}}
\providecommand{\dodoi}[1]{doi:~\href{http://doi.org/#1}{\nolinkurl{#1}}}
\providecommand{\doeprint}[1]{\href{http://ascl.net/#1}{\nolinkurl{http://ascl.net/#1}}}
\providecommand{\doarXiv}[1]{\href{https://arxiv.org/abs/#1}{\nolinkurl{https://arxiv.org/abs/#1}}}

\bibitem[{{Alonso} {et~al.}(2008){Alonso}, {Auvergne}, {Baglin}, {Ollivier},
  {Moutou}, {Rouan}, {Deeg}, {Aigrain}, {Almenara}, {Barbieri}, {Barge},
  {Benz}, {Bord{\'e}}, {Bouchy}, {de La Reza}, {Deleuil}, {Dvorak}, {Erikson},
  {Fridlund}, {Gillon}, {Gondoin}, {Guillot}, {Hatzes}, {H{\'e}brard},
  {Kabath}, {Jorda}, {Lammer}, {L{\'e}ger}, {Llebaria}, {Loeillet}, {Magain},
  {Mayor}, {Mazeh}, {P{\"a}tzold}, {Pepe}, {Pont}, {Queloz}, {Rauer},
  {Shporer}, {Schneider}, {Stecklum}, {Udry}, \& {Wuchterl}}]{alonso2008}
{Alonso}, R., {Auvergne}, M., {Baglin}, A., {et~al.} 2008, \aap, 482, L21,
  \dodoi{10.1051/0004-6361:200809431}

\bibitem[{{Amard} {et~al.}(2016){Amard}, {Palacios}, {Charbonnel}, {Gallet}, \&
  {Bouvier}}]{amard2016}
{Amard}, L., {Palacios}, A., {Charbonnel}, C., {Gallet}, F., \& {Bouvier}, J.
  2016, \aap, 587, A105, \dodoi{10.1051/0004-6361/201527349}

\bibitem[{{Anderson} {et~al.}(2012){Anderson}, {Collier Cameron}, {Gillon},
  {Hellier}, {Jehin}, {Lendl}, {Maxted}, {Queloz}, {Smalley}, {Smith},
  {Triaud}, {West}, {Pepe}, {Pollacco}, {S{\'e}gransan}, {Todd}, \&
  {Udry}}]{anderson2012}
{Anderson}, D.~R., {Collier Cameron}, A., {Gillon}, M., {et~al.} 2012, \mnras,
  422, 1988, \dodoi{10.1111/j.1365-2966.2012.20635.x}

\bibitem[{{Anderson} {et~al.}(2014){Anderson}, {Collier Cameron}, {Delrez},
  {Doyle}, {Faedi}, {Fumel}, {Gillon}, {G{\'o}mez Maqueo Chew}, {Hellier},
  {Jehin}, {Lendl}, {Maxted}, {Pepe}, {Pollacco}, {Queloz}, {S{\'e}gransan},
  {Skillen}, {Smalley}, {Smith}, {Southworth}, {Triaud}, {Turner}, {Udry}, \&
  {West}}]{anderson2014}
{Anderson}, D.~R., {Collier Cameron}, A., {Delrez}, L., {et~al.} 2014, \mnras,
  445, 1114, \dodoi{10.1093/mnras/stu1737}

\bibitem[{{Anderson} {et~al.}(2016){Anderson}, {Storch}, \&
  {Lai}}]{anderson2016}
{Anderson}, K.~R., {Storch}, N.~I., \& {Lai}, D. 2016, \mnras, 456, 3671,
  \dodoi{10.1093/mnras/stv2906}

\bibitem[{{Bailey} \& {Goodman}(2019)}]{BaileyGoodman2019}
{Bailey}, A., \& {Goodman}, J. 2019, \mnras, 482, 1872,
  \dodoi{10.1093/mnras/sty2805}

\bibitem[{{Bakos} {et~al.}(2007){Bakos}, {Shporer}, {P{\'a}l}, {Torres},
  {Kov{\'a}cs}, {Latham}, {Mazeh}, {Ofir}, {Noyes}, {Sasselov}, {Bouchy},
  {Pont}, {Queloz}, {Udry}, {Esquerdo}, {Sip{\H{o}}cz}, {Kov{\'a}cs},
  {Stefanik}, {L{\'a}z{\'a}r}, {Papp}, \& {S{\'a}ri}}]{bakos2007}
{Bakos}, G.~{\'A}., {Shporer}, A., {P{\'a}l}, A., {et~al.} 2007, \apjl, 671,
  L173, \dodoi{10.1086/525022}

\bibitem[{{Bakos} {et~al.}(2009){Bakos}, {Howard}, {Noyes}, {Hartman},
  {Torres}, {Kov{\'a}cs}, {Fischer}, {Latham}, {Johnson}, {Marcy}, {Sasselov},
  {Stefanik}, {Sip{\H{o}}cz}, {Kov{\'a}cs}, {Esquerdo}, {P{\'a}l},
  {L{\'a}z{\'a}r}, {Papp}, \& {S{\'a}ri}}]{bakos2009}
{Bakos}, G.~{\'A}., {Howard}, A.~W., {Noyes}, R.~W., {et~al.} 2009, \apj, 707,
  446, \dodoi{10.1088/0004-637X/707/1/446}

\bibitem[{{Bakos} {et~al.}(2012){Bakos}, {Hartman}, {Torres}, {B{\'e}ky},
  {Latham}, {Buchhave}, {Csubry}, {Kov{\'a}cs}, {Bieryla}, {Quinn},
  {Szklen{\'a}r}, {Esquerdo}, {Shporer}, {Noyes}, {Fischer}, {Johnson},
  {Howard}, {Marcy}, {Sato}, {Penev}, {Everett}, {Sasselov},
  {F{\H{u}}r{\'e}sz}, {Stefanik}, {L{\'a}z{\'a}r}, {Papp}, \&
  {S{\'a}ri}}]{bakos2012}
{Bakos}, G.~{\'A}., {Hartman}, J.~D., {Torres}, G., {et~al.} 2012, \aj, 144,
  19, \dodoi{10.1088/0004-6256/144/1/19}

\bibitem[{{Barker}(2020)}]{barker2020}
{Barker}, A.~J. 2020, \mnras, 498, 2270, \dodoi{10.1093/mnras/staa2405}

\bibitem[{{Barker} \& {Ogilvie}(2009)}]{barker2009}
{Barker}, A.~J., \& {Ogilvie}, G.~I. 2009, \mnras, 395, 2268,
  \dodoi{10.1111/j.1365-2966.2009.14694.x}

\bibitem[{{Barker} \& {Ogilvie}(2010)}]{barker2010}
---. 2010, \mnras, 404, 1849, \dodoi{10.1111/j.1365-2966.2010.16400.x}

\bibitem[{{Barros} {et~al.}(2016){Barros}, {Brown}, {H{\'e}brard}, {G{\'o}mez
  Maqueo Chew}, {Anderson}, {Boumis}, {Delrez}, {Hay}, {Lam}, {Llama}, {Lendl},
  {McCormac}, {Skiff}, {Smalley}, {Turner}, {Vanhuysse}, {Armstrong}, {Boisse},
  {Bouchy}, {Collier Cameron}, {Faedi}, {Gillon}, {Hellier}, {Jehin}, {Liakos},
  {Meaburn}, {Osborn}, {Pepe}, {Plauchu-Frayn}, {Pollacco}, {Queloz}, {Rey},
  {Spake}, {S{\'e}gransan}, {Triaud}, {Udry}, {Walker}, {Watson}, {West}, \&
  {Wheatley}}]{barros2016}
{Barros}, S.~C.~C., {Brown}, D.~J.~A., {H{\'e}brard}, G., {et~al.} 2016, \aap,
  593, A113, \dodoi{10.1051/0004-6361/201526517}

\bibitem[{{Bate} {et~al.}(2010){Bate}, {Lodato}, \& {Pringle}}]{bate2010}
{Bate}, M.~R., {Lodato}, G., \& {Pringle}, J.~E. 2010, \mnras, 401, 1505,
  \dodoi{10.1111/j.1365-2966.2009.15773.x}

\bibitem[{{Batygin}(2012)}]{batygin2012}
{Batygin}, K. 2012, \nat, 491, 418, \dodoi{10.1038/nature11560}

\bibitem[{{Batygin} {et~al.}(2016){Batygin}, {Bodenheimer}, \&
  {Laughlin}}]{batygin2016}
{Batygin}, K., {Bodenheimer}, P.~H., \& {Laughlin}, G.~P. 2016, \apj, 829, 114,
  \dodoi{10.3847/0004-637X/829/2/114}

\bibitem[{{Beaug{\'e}} \& {Nesvorn{\'y}}(2012)}]{beauge2012}
{Beaug{\'e}}, C., \& {Nesvorn{\'y}}, D. 2012, \apj, 751, 119,
  \dodoi{10.1088/0004-637X/751/2/119}

\bibitem[{{Boisse} {et~al.}(2013){Boisse}, {Hartman}, {Bakos}, {Penev},
  {Csubry}, {B{\'e}ky}, {Latham}, {Bieryla}, {Torres}, {Kov{\'a}cs},
  {Buchhave}, {Hansen}, {Everett}, {Esquerdo}, {Szklen{\'a}r}, {Falco},
  {Shporer}, {Fulton}, {Noyes}, {Stefanik}, {L{\'a}z{\'a}r}, {Papp}, \&
  {S{\'a}ri}}]{boisse2013}
{Boisse}, I., {Hartman}, J.~D., {Bakos}, G.~{\'A}., {et~al.} 2013, \aap, 558,
  A86, \dodoi{10.1051/0004-6361/201220993}

\bibitem[{{Boley} {et~al.}(2016){Boley}, {Granados Contreras}, \&
  {Gladman}}]{boley2016}
{Boley}, A.~C., {Granados Contreras}, A.~P., \& {Gladman}, B. 2016, \apjl, 817,
  L17, \dodoi{10.3847/2041-8205/817/2/L17}

\bibitem[{{Bolmont} \& {Mathis}(2016)}]{bolmont2016}
{Bolmont}, E., \& {Mathis}, S. 2016, Celestial Mechanics and Dynamical
  Astronomy, 126, 275, \dodoi{10.1007/s10569-016-9690-3}

\bibitem[{{Bonomo} {et~al.}(2015){Bonomo}, {Sozzetti}, {Santerne}, {Deleuil},
  {Almenara}, {Bruno}, {D{\'\i}az}, {H{\'e}brard}, \& {Moutou}}]{bonomo2015}
{Bonomo}, A.~S., {Sozzetti}, A., {Santerne}, A., {et~al.} 2015, \aap, 575, A85,
  \dodoi{10.1051/0004-6361/201323042}

\bibitem[{{Borucki} {et~al.}(2011){Borucki}, {Koch}, {Basri}, {Batalha},
  {Brown}, {Bryson}, {Caldwell}, {Christensen-Dalsgaard}, {Cochran}, {DeVore},
  {Dunham}, {Gautier}, {Geary}, {Gilliland}, {Gould}, {Howell}, {Jenkins},
  {Latham}, {Lissauer}, {Marcy}, {Rowe}, {Sasselov}, {Boss}, {Charbonneau},
  {Ciardi}, {Doyle}, {Dupree}, {Ford}, {Fortney}, {Holman}, {Seager},
  {Steffen}, {Tarter}, {Welsh}, {Allen}, {Buchhave}, {Christiansen}, {Clarke},
  {Das}, {D{\'e}sert}, {Endl}, {Fabrycky}, {Fressin}, {Haas}, {Horch},
  {Howard}, {Isaacson}, {Kjeldsen}, {Kolodziejczak}, {Kulesa}, {Li}, {Lucas},
  {Machalek}, {McCarthy}, {MacQueen}, {Meibom}, {Miquel}, {Prsa}, {Quinn},
  {Quintana}, {Ragozzine}, {Sherry}, {Shporer}, {Tenenbaum}, {Torres},
  {Twicken}, {Van Cleve}, {Walkowicz}, {Witteborn}, \& {Still}}]{borucki2011}
{Borucki}, W.~J., {Koch}, D.~G., {Basri}, G., {et~al.} 2011, \apj, 736, 19,
  \dodoi{10.1088/0004-637X/736/1/19}

\bibitem[{{Bouchy} {et~al.}(2008){Bouchy}, {Queloz}, {Deleuil}, {Loeillet},
  {Hatzes}, {Aigrain}, {Alonso}, {Auvergne}, {Baglin}, {Barge}, {Benz},
  {Bord{\'e}}, {Deeg}, {de La Reza}, {Dvorak}, {Erikson}, {Fridlund},
  {Gondoin}, {Guillot}, {H{\'e}brard}, {Jorda}, {Lammer}, {L{\'e}ger},
  {Llebaria}, {Magain}, {Mayor}, {Moutou}, {Ollivier}, {P{\"a}tzold}, {Pepe},
  {Pont}, {Rauer}, {Rouan}, {Schneider}, {Triaud}, {Udry}, \&
  {Wuchterl}}]{bouchy2008}
{Bouchy}, F., {Queloz}, D., {Deleuil}, M., {et~al.} 2008, \aap, 482, L25,
  \dodoi{10.1051/0004-6361:200809433}

\bibitem[{{Bovy} {et~al.}(2016){Bovy}, {Rix}, {Green}, {Schlafly}, \&
  {Finkbeiner}}]{bovy2016}
{Bovy}, J., {Rix}, H.-W., {Green}, G.~M., {Schlafly}, E.~F., \& {Finkbeiner},
  D.~P. 2016, \apj, 818, 130, \dodoi{10.3847/0004-637X/818/2/130}

\bibitem[{{Brown} {et~al.}(2012){Brown}, {Cameron}, {Anderson}, {Enoch},
  {Hellier}, {Maxted}, {Miller}, {Pollacco}, {Queloz}, {Simpson}, {Smalley},
  {Triaud}, {Boisse}, {Bouchy}, {Gillon}, \& {H{\'e}brard}}]{brown2012}
{Brown}, D.~J.~A., {Cameron}, A.~C., {Anderson}, D.~R., {et~al.} 2012, \mnras,
  423, 1503, \dodoi{10.1111/j.1365-2966.2012.20973.x}

\bibitem[{{Cabrera} {et~al.}(2010){Cabrera}, {Bruntt}, {Ollivier}, {D{\'\i}az},
  {Csizmadia}, {Aigrain}, {Alonso}, {Almenara}, {Auvergne}, {Baglin}, {Barge},
  {Bonomo}, {Bord{\'e}}, {Bouchy}, {Carone}, {Carpano}, {Deleuil}, {Deeg},
  {Dvorak}, {Erikson}, {Ferraz-Mello}, {Fridlund}, {Gandolfi}, {Gazzano},
  {Gillon}, {Guenther}, {Guillot}, {Hatzes}, {Havel}, {H{\'e}brard}, {Jorda},
  {L{\'e}ger}, {Llebaria}, {Lammer}, {Lovis}, {Mazeh}, {Moutou}, {Ofir}, {von
  Paris}, {P{\"a}tzold}, {Queloz}, {Rauer}, {Rouan}, {Santerne}, {Schneider},
  {Tingley}, {Titz-Weider}, \& {Wuchterl}}]{cabrera2010}
{Cabrera}, J., {Bruntt}, H., {Ollivier}, M., {et~al.} 2010, \aap, 522, A110,
  \dodoi{10.1051/0004-6361/201015154}

\bibitem[{{Choi} {et~al.}(2016){Choi}, {Dotter}, {Conroy}, {Cantiello},
  {Paxton}, \& {Johnson}}]{choi2016}
{Choi}, J., {Dotter}, A., {Conroy}, C., {et~al.} 2016, \apj, 823, 102,
  \dodoi{10.3847/0004-637X/823/2/102}

\bibitem[{{Ciceri} {et~al.}(2016){Ciceri}, {Mancini}, {Henning}, {Bakos},
  {Penev}, {Brahm}, {Zhou}, {Hartman}, {Bayliss}, {Jord{\'a}n}, {Csubry}, {de
  Val-Borro}, {Bhatti}, {Rabus}, {Espinoza}, {Suc}, {Schmidt}, {Noyes},
  {Howard}, {Fulton}, {Isaacson}, {Marcy}, {Butler}, {Arriagada}, {Crane},
  {Shectman}, {Thompson}, {Tan}, {L{\'a}z{\'a}r}, {Papp}, \&
  {Sari}}]{ciceri2016}
{Ciceri}, S., {Mancini}, L., {Henning}, T., {et~al.} 2016, \pasp, 128, 074401,
  \dodoi{10.1088/1538-3873/128/965/074401}

\bibitem[{{Collier Cameron} \& {Jardine}(2018)}]{collier_cameron2018}
{Collier Cameron}, A., \& {Jardine}, M. 2018, \mnras, 476, 2542,
  \dodoi{10.1093/mnras/sty292}

\bibitem[{{Csizmadia} {et~al.}(2011){Csizmadia}, {Moutou}, {Deleuil},
  {Cabrera}, {Fridlund}, {Gandolfi}, {Aigrain}, {Alonso}, {Almenara},
  {Auvergne}, {Baglin}, {Barge}, {Bonomo}, {Bord{\'e}}, {Bouchy}, {Bruntt},
  {Carone}, {Carpano}, {Cavarroc}, {Cochran}, {Deeg}, {D{\'\i}az}, {Dvorak},
  {Endl}, {Erikson}, {Ferraz-Mello}, {Fruth}, {Gazzano}, {Gillon}, {Guenther},
  {Guillot}, {Hatzes}, {Havel}, {H{\'e}brard}, {Jehin}, {Jorda}, {L{\'e}ger},
  {Llebaria}, {Lammer}, {Lovis}, {MacQueen}, {Mazeh}, {Ollivier},
  {P{\"a}tzold}, {Queloz}, {Rauer}, {Rouan}, {Santerne}, {Schneider},
  {Tingley}, {Titz-Weider}, \& {Wuchterl}}]{csizmadia2011}
{Csizmadia}, S., {Moutou}, C., {Deleuil}, M., {et~al.} 2011, \aap, 531, A41,
  \dodoi{10.1051/0004-6361/201117009}

\bibitem[{{Damiani} \& {Lanza}(2015)}]{damiani2015}
{Damiani}, C., \& {Lanza}, A.~F. 2015, \aap, 574, A39,
  \dodoi{10.1051/0004-6361/201424318}

\bibitem[{{Dawson} \& {Johnson}(2018)}]{dawson2018}
{Dawson}, R.~I., \& {Johnson}, J.~A. 2018, \araa, 56, 175,
  \dodoi{10.1146/annurev-astro-081817-051853}

\bibitem[{{Deleuil} {et~al.}(2014){Deleuil}, {Almenara}, {Santerne}, {Barros},
  {Havel}, {H{\'e}brard}, {Bonomo}, {Bouchy}, {Bruno}, {Damiani}, {D{\'\i}az},
  {Montagnier}, \& {Moutou}}]{deleuil2014}
{Deleuil}, M., {Almenara}, J.~M., {Santerne}, A., {et~al.} 2014, \aap, 564,
  A56, \dodoi{10.1051/0004-6361/201323017}

\bibitem[{{D{\'e}sert} {et~al.}(2011){D{\'e}sert}, {Charbonneau}, {Demory},
  {Ballard}, {Carter}, {Fortney}, {Cochran}, {Endl}, {Quinn}, {Isaacson},
  {Fressin}, {Buchhave}, {Latham}, {Knutson}, {Bryson}, {Torres}, {Rowe},
  {Batalha}, {Borucki}, {Brown}, {Caldwell}, {Christiansen}, {Deming},
  {Fabrycky}, {Ford}, {Gilliland}, {Gillon}, {Haas}, {Jenkins}, {Kinemuchi},
  {Koch}, {Lissauer}, {Lucas}, {Mullally}, {MacQueen}, {Marcy}, {Sasselov},
  {Seager}, {Still}, {Tenenbaum}, {Uddin}, \& {Winn}}]{desert2011}
{D{\'e}sert}, J.-M., {Charbonneau}, D., {Demory}, B.-O., {et~al.} 2011, \apjs,
  197, 14, \dodoi{10.1088/0067-0049/197/1/14}

\bibitem[{{Dotter}(2016)}]{dotter2016}
{Dotter}, A. 2016, \apjs, 222, 8, \dodoi{10.3847/0067-0049/222/1/8}

\bibitem[{{Duguid} {et~al.}(2020){Duguid}, {Barker}, \& {Jones}}]{duguid2020}
{Duguid}, C.~D., {Barker}, A.~J., \& {Jones}, C.~A. 2020, \mnras, 497, 3400,
  \dodoi{10.1093/mnras/staa2216}

\bibitem[{{Dunham} {et~al.}(2010){Dunham}, {Borucki}, {Koch}, {Batalha},
  {Buchhave}, {Brown}, {Caldwell}, {Cochran}, {Endl}, {Fischer},
  {F{\H{u}}r{\'e}sz}, {Gautier}, {Geary}, {Gilliland}, {Gould}, {Howell},
  {Jenkins}, {Kjeldsen}, {Latham}, {Lissauer}, {Marcy}, {Meibom}, {Monet},
  {Rowe}, \& {Sasselov}}]{dunham2010}
{Dunham}, E.~W., {Borucki}, W.~J., {Koch}, D.~G., {et~al.} 2010, \apjl, 713,
  L136, \dodoi{10.1088/2041-8205/713/2/L136}

\bibitem[{{Endl} {et~al.}(2014){Endl}, {Caldwell}, {Barclay}, {Huber},
  {Isaacson}, {Buchhave}, {Brugamyer}, {Robertson}, {Cochran}, {MacQueen},
  {Havel}, {Lucas}, {Howell}, {Fischer}, {Quintana}, \& {Ciardi}}]{endl2014}
{Endl}, M., {Caldwell}, D.~A., {Barclay}, T., {et~al.} 2014, \apj, 795, 151,
  \dodoi{10.1088/0004-637X/795/2/151}

\bibitem[{{Espinoza} {et~al.}(2019){Espinoza}, {Hartman}, {Bakos}, {Henning},
  {Bayliss}, {Bento}, {Bhatti}, {Brahm}, {Csubry}, {Suc}, {Jord{\'a}n},
  {Mancini}, {Tan}, {Penev}, {Rabus}, {Sarkis}, {de Val-Borro}, {Durkan},
  {L{\'a}z{\'a}r}, {Papp}, \& {S{\'a}ri}}]{espinoza2019}
{Espinoza}, N., {Hartman}, J.~D., {Bakos}, G.~{\'A}., {et~al.} 2019, \aj, 158,
  63, \dodoi{10.3847/1538-3881/ab26bb}

\bibitem[{{Evans} {et~al.}(2018){Evans}, {Riello}, {De Angeli}, {Carrasco},
  {Montegriffo}, {Fabricius}, {Jordi}, {Palaversa}, {Diener}, {Busso},
  {Cacciari}, {van Leeuwen}, {Burgess}, {Davidson}, {Harrison}, {Hodgkin},
  {Pancino}, {Richards}, {Altavilla}, {Balaguer-N{\'u}{\~n}ez}, {Barstow},
  {Bellazzini}, {Brown}, {Castellani}, {Cocozza}, {De Luise}, {Delgado},
  {Ducourant}, {Galleti}, {Gilmore}, {Giuffrida}, {Holl}, {Kewley}, {Koposov},
  {Marinoni}, {Marrese}, {Osborne}, {Piersimoni}, {Portell}, {Pulone},
  {Ragaini}, {Sanna}, {Terrett}, {Walton}, {Wevers}, \&
  {Wyrzykowski}}]{evans2018}
{Evans}, D.~W., {Riello}, M., {De Angeli}, F., {et~al.} 2018, \aap, 616, A4,
  \dodoi{10.1051/0004-6361/201832756}

\bibitem[{{Fabrycky} \& {Tremaine}(2007)}]{fabrycky2007}
{Fabrycky}, D., \& {Tremaine}, S. 2007, \apj, 669, 1298, \dodoi{10.1086/521702}

\bibitem[{{Faedi} {et~al.}(2013){Faedi}, {Pollacco}, {Barros}, {Brown},
  {Collier Cameron}, {Doyle}, {Enoch}, {Gillon}, {G{\'o}mez Maqueo Chew},
  {H{\'e}brard}, {Lendl}, {Liebig}, {Smalley}, {Triaud}, {West}, {Wheatley},
  {Alsubai}, {Anderson}, {Armstrong}, {Bento}, {Bochinski}, {Bouchy},
  {Busuttil}, {Fossati}, {Fumel}, {Haswell}, {Hellier}, {Holmes}, {Jehin},
  {Kolb}, {McCormac}, {Miller}, {Moutou}, {Norton}, {Parley}, {Queloz},
  {Santerne}, {Skillen}, {Smith}, {Udry}, \& {Watson}}]{faedi2013}
{Faedi}, F., {Pollacco}, D., {Barros}, S.~C.~C., {et~al.} 2013, \aap, 551, A73,
  \dodoi{10.1051/0004-6361/201220520}

\bibitem[{{Fulton} {et~al.}(2015){Fulton}, {Collins}, {Gaudi}, {Stassun},
  {Pepper}, {Beatty}, {Siverd}, {Penev}, {Howard}, {Baranec}, {Corfini},
  {Eastman}, {Gregorio}, {Law}, {Lund}, {Oberst}, {Penny}, {Riddle},
  {Rodriguez}, {Stevens}, {Zambelli}, {Ziegler}, {Bieryla}, {D'Ago}, {DePoy},
  {Jensen}, {Kielkopf}, {Latham}, {Manner}, {Marshall}, {McLeod}, \&
  {Reed}}]{fulton2015}
{Fulton}, B.~J., {Collins}, K.~A., {Gaudi}, B.~S., {et~al.} 2015, \apj, 810,
  30, \dodoi{10.1088/0004-637X/810/1/30}

\bibitem[{{Gaia Collaboration} {et~al.}(2018){Gaia Collaboration}, {Brown},
  {Vallenari}, {Prusti}, {de Bruijne}, {Babusiaux}, {Bailer-Jones}, {Biermann},
  {Evans}, {Eyer}, {Jansen}, {Jordi}, {Klioner}, {Lammers}, {Lindegren},
  {Luri}, {Mignard}, {Panem}, {Pourbaix}, {Randich}, {Sartoretti}, {Siddiqui},
  {Soubiran}, {van Leeuwen}, {Walton}, {Arenou}, {Bastian}, {Cropper},
  {Drimmel}, {Katz}, {Lattanzi}, {Bakker}, {Cacciari}, {Casta{\~n}eda},
  {Chaoul}, {Cheek}, {De Angeli}, {Fabricius}, {Guerra}, {Holl}, {Masana},
  {Messineo}, {Mowlavi}, {Nienartowicz}, {Panuzzo}, {Portell}, {Riello},
  {Seabroke}, {Tanga}, {Th{\'e}venin}, {Gracia-Abril}, {Comoretto},
  {Garcia-Reinaldos}, {Teyssier}, {Altmann}, {Andrae}, {Audard},
  {Bellas-Velidis}, {Benson}, {Berthier}, {Blomme}, {Burgess}, {Busso},
  {Carry}, {Cellino}, {Clementini}, {Clotet}, {Creevey}, {Davidson}, {De
  Ridder}, {Delchambre}, {Dell'Oro}, {Ducourant},
  {Fern{\'a}ndez-Hern{\'a}ndez}, {Fouesneau}, {Fr{\'e}mat}, {Galluccio},
  {Garc{\'\i}a-Torres}, {Gonz{\'a}lez-N{\'u}{\~n}ez}, {Gonz{\'a}lez-Vidal},
  {Gosset}, {Guy}, {Halbwachs}, {Hambly}, {Harrison}, {Hern{\'a}ndez},
  {Hestroffer}, {Hodgkin}, {Hutton}, {Jasniewicz}, {Jean-Antoine-Piccolo},
  {Jordan}, {Korn}, {Krone-Martins}, {Lanzafame}, {Lebzelter}, {L{\"o}ffler},
  {Manteiga}, {Marrese}, {Mart{\'\i}n-Fleitas}, {Moitinho}, {Mora}, {Muinonen},
  {Osinde}, {Pancino}, {Pauwels}, {Petit}, {Recio-Blanco}, {Richards},
  {Rimoldini}, {Robin}, {Sarro}, {Siopis}, {Smith}, {Sozzetti}, {S{\"u}veges},
  {Torra}, {van Reeven}, {Abbas}, {Abreu Aramburu}, {Accart}, {Aerts},
  {Altavilla}, {{\'A}lvarez}, {Alvarez}, {Alves}, {Anderson}, {Andrei},
  {Anglada Varela}, {Antiche}, {Antoja}, {Arcay}, {Astraatmadja}, {Bach},
  {Baker}, {Balaguer-N{\'u}{\~n}ez}, {Balm}, {Barache}, {Barata}, {Barbato},
  {Barblan}, {Barklem}, {Barrado}, {Barros}, {Barstow}, {Bartholom{\'e}
  Mu{\~n}oz}, {Bassilana}, {Becciani}, {Bellazzini}, {Berihuete}, {Bertone},
  {Bianchi}, {Bienaym{\'e}}, {Blanco-Cuaresma}, {Boch}, {Boeche}, {Bombrun},
  {Borrachero}, {Bossini}, {Bouquillon}, {Bourda}, {Bragaglia}, {Bramante},
  {Breddels}, {Bressan}, {Brouillet}, {Br{\"u}semeister}, {Brugaletta},
  {Bucciarelli}, {Burlacu}, {Busonero}, {Butkevich}, {Buzzi}, {Caffau},
  {Cancelliere}, {Cannizzaro}, {Cantat-Gaudin}, {Carballo}, {Carlucci},
  {Carrasco}, {Casamiquela}, {Castellani}, {Castro-Ginard}, {Charlot},
  {Chemin}, {Chiavassa}, {Cocozza}, {Costigan}, {Cowell}, {Crifo}, {Crosta},
  {Crowley}, {Cuypers}, {Dafonte}, {Damerdji}, {Dapergolas}, {David}, {David},
  {de Laverny}, {De Luise}, {De March}, {de Martino}, {de Souza}, {de Torres},
  {Debosscher}, {del Pozo}, {Delbo}, {Delgado}, {Delgado}, {Di Matteo},
  {Diakite}, {Diener}, {Distefano}, {Dolding}, {Drazinos}, {Dur{\'a}n},
  {Edvardsson}, {Enke}, {Eriksson}, {Esquej}, {Eynard Bontemps}, {Fabre},
  {Fabrizio}, {Faigler}, {Falc{\~a}o}, {Farr{\`a}s Casas}, {Federici},
  {Fedorets}, {Fernique}, {Figueras}, {Filippi}, {Findeisen}, {Fonti},
  {Fraile}, {Fraser}, {Fr{\'e}zouls}, {Gai}, {Galleti}, {Garabato},
  {Garc{\'\i}a-Sedano}, {Garofalo}, {Garralda}, {Gavel}, {Gavras}, {Gerssen},
  {Geyer}, {Giacobbe}, {Gilmore}, {Girona}, {Giuffrida}, {Glass}, {Gomes},
  {Granvik}, {Gueguen}, {Guerrier}, {Guiraud}, {Guti{\'e}rrez-S{\'a}nchez},
  {Haigron}, {Hatzidimitriou}, {Hauser}, {Haywood}, {Heiter}, {Helmi}, {Heu},
  {Hilger}, {Hobbs}, {Hofmann}, {Holland}, {Huckle}, {Hypki}, {Icardi},
  {Jan{\ss}en}, {Jevardat de Fombelle}, {Jonker}, {Juh{\'a}sz}, {Julbe},
  {Karampelas}, {Kewley}, {Klar}, {Kochoska}, {Kohley}, {Kolenberg},
  {Kontizas}, {Kontizas}, {Koposov}, {Kordopatis}, {Kostrzewa-Rutkowska},
  {Koubsky}, {Lambert}, {Lanza}, {Lasne}, {Lavigne}, {Le Fustec}, {Le
  Poncin-Lafitte}, {Lebreton}, {Leccia}, {Leclerc}, {Lecoeur-Taibi},
  {Lenhardt}, {Leroux}, {Liao}, {Licata}, {Lindstr{\o}m}, {Lister}, {Livanou},
  {Lobel}, {L{\'o}pez}, {Managau}, {Mann}, {Mantelet}, {Marchal}, {Marchant},
  {Marconi}, {Marinoni}, {Marschalk{\'o}}, {Marshall}, {Martino}, {Marton},
  {Mary}, {Massari}, {Matijevi{\v{c}}}, {Mazeh}, {McMillan}, {Messina},
  {Michalik}, {Millar}, {Molina}, {Molinaro}, {Moln{\'a}r}, {Montegriffo},
  {Mor}, {Morbidelli}, {Morel}, {Morris}, {Mulone}, {Muraveva}, {Musella},
  {Nelemans}, {Nicastro}, {Noval}, {O'Mullane}, {Ord{\'e}novic},
  {Ord{\'o}{\~n}ez-Blanco}, {Osborne}, {Pagani}, {Pagano}, {Pailler},
  {Palacin}, {Palaversa}, {Panahi}, {Pawlak}, {Piersimoni}, {Pineau}, {Plachy},
  {Plum}, {Poggio}, {Poujoulet}, {Pr{\v{s}}a}, {Pulone}, {Racero}, {Ragaini},
  {Rambaux}, {Ramos-Lerate}, {Regibo}, {Reyl{\'e}}, {Riclet}, {Ripepi}, {Riva},
  {Rivard}, {Rixon}, {Roegiers}, {Roelens}, {Romero-G{\'o}mez}, {Rowell},
  {Royer}, {Ruiz-Dern}, {Sadowski}, {Sagrist{\`a} Sell{\'e}s}, {Sahlmann},
  {Salgado}, {Salguero}, {Sanna}, {Santana-Ros}, {Sarasso}, {Savietto},
  {Schultheis}, {Sciacca}, {Segol}, {Segovia}, {S{\'e}gransan}, {Shih},
  {Siltala}, {Silva}, {Smart}, {Smith}, {Solano}, {Solitro}, {Sordo}, {Soria
  Nieto}, {Souchay}, {Spagna}, {Spoto}, {Stampa}, {Steele},
  {Steidelm{\"u}ller}, {Stephenson}, {Stoev}, {Suess}, {Surdej}, {Szabados},
  {Szegedi-Elek}, {Tapiador}, {Taris}, {Tauran}, {Taylor}, {Teixeira},
  {Terrett}, {Teyssand ier}, {Thuillot}, {Titarenko}, {Torra Clotet}, {Turon},
  {Ulla}, {Utrilla}, {Uzzi}, {Vaillant}, {Valentini}, {Valette}, {van Elteren},
  {Van Hemelryck}, {van Leeuwen}, {Vaschetto}, {Vecchiato}, {Veljanoski},
  {Viala}, {Vicente}, {Vogt}, {von Essen}, {Voss}, {Votruba}, {Voutsinas},
  {Walmsley}, {Weiler}, {Wertz}, {Wevers}, {Wyrzykowski}, {Yoldas},
  {{\v{Z}}erjal}, {Ziaeepour}, {Zorec}, {Zschocke}, {Zucker}, {Zurbach}, \&
  {Zwitter}}]{gaia2018}
{Gaia Collaboration}, {Brown}, A.~G.~A., {Vallenari}, A., {et~al.} 2018, \aap,
  616, A1, \dodoi{10.1051/0004-6361/201833051}

\bibitem[{{Gallet} \& {Bouvier}(2013)}]{gallet2013}
{Gallet}, F., \& {Bouvier}, J. 2013, \aap, 556, A36,
  \dodoi{10.1051/0004-6361/201321302}

\bibitem[{{Gillon} {et~al.}(2011){Gillon}, {Doyle}, {Lendl}, {Maxted},
  {Triaud}, {Anderson}, {Barros}, {Bento}, {Collier-Cameron}, {Enoch}, {Faedi},
  {Hellier}, {Jehin}, {Magain}, {Montalb{\'a}n}, {Pepe}, {Pollacco}, {Queloz},
  {Smalley}, {Segransan}, {Smith}, {Southworth}, {Udry}, {West}, \&
  {Wheatley}}]{gillon2011}
{Gillon}, M., {Doyle}, A.~P., {Lendl}, M., {et~al.} 2011, \aap, 533, A88,
  \dodoi{10.1051/0004-6361/201117198}

\bibitem[{{Gillon} {et~al.}(2013){Gillon}, {Anderson}, {Collier-Cameron},
  {Doyle}, {Fumel}, {Hellier}, {Jehin}, {Lendl}, {Maxted}, {Montalb{\'a}n},
  {Pepe}, {Pollacco}, {Queloz}, {S{\'e}gransan}, {Smith}, {Smalley},
  {Southworth}, {Triaud}, {Udry}, \& {West}}]{gillon2013}
{Gillon}, M., {Anderson}, D.~R., {Collier-Cameron}, A., {et~al.} 2013, \aap,
  552, A82, \dodoi{10.1051/0004-6361/201220561}

\bibitem[{{Goldreich} \& {Nicholson}(1977)}]{goldreich1977}
{Goldreich}, P., \& {Nicholson}, P.~D. 1977, \icarus, 30, 301,
  \dodoi{10.1016/0019-1035(77)90163-4}

\bibitem[{{G{\'o}mez Maqueo Chew} {et~al.}(2013){G{\'o}mez Maqueo Chew},
  {Faedi}, {Pollacco}, {Brown}, {Doyle}, {Collier Cameron}, {Gillon}, {Lendl},
  {Smalley}, {Triaud}, {West}, {Wheatley}, {Busuttil}, {Liebig}, {Anderson},
  {Armstrong}, {Barros}, {Bento}, {Bochinski}, {Burwitz}, {Delrez}, {Enoch},
  {Fumel}, {Haswell}, {H{\'e}brard}, {Hellier}, {Holmes}, {Jehin}, {Kolb},
  {Maxted}, {McCormac}, {Miller}, {Norton}, {Pepe}, {Queloz}, {Rodr{\'\i}guez},
  {S{\'e}gransan}, {Skillen}, {Stassun}, {Udry}, \&
  {Watson}}]{gomez-maqueo-chew2013}
{G{\'o}mez Maqueo Chew}, Y., {Faedi}, F., {Pollacco}, D., {et~al.} 2013, \aap,
  559, A36, \dodoi{10.1051/0004-6361/201322314}

\bibitem[{{Goodman} \& {Dickson}(1998)}]{goodman1998}
{Goodman}, J., \& {Dickson}, E.~S. 1998, \apj, 507, 938, \dodoi{10.1086/306348}

\bibitem[{{Goodman} \& {Lackner}(2009)}]{goodman2009}
{Goodman}, J., \& {Lackner}, C. 2009, \apj, 696, 2054,
  \dodoi{10.1088/0004-637X/696/2/2054}

\bibitem[{{Goodman} \& {Oh}(1997)}]{goodman1997}
{Goodman}, J., \& {Oh}, S.~P. 1997, \apj, 486, 403, \dodoi{10.1086/304505}

\bibitem[{{Guillochon} {et~al.}(2011){Guillochon}, {Ramirez-Ruiz}, \&
  {Lin}}]{guillochon2011}
{Guillochon}, J., {Ramirez-Ruiz}, E., \& {Lin}, D. 2011, \apj, 732, 74,
  \dodoi{10.1088/0004-637X/732/2/74}

\bibitem[{{Hamer} \& {Schlaufman}(2019)}]{hamer2019}
{Hamer}, J.~H., \& {Schlaufman}, K.~C. 2019, \aj, 158, 190,
  \dodoi{10.3847/1538-3881/ab3c56}

\bibitem[{{Hebb} {et~al.}(2010){Hebb}, {Collier-Cameron}, {Triaud}, {Lister},
  {Smalley}, {Maxted}, {Hellier}, {Anderson}, {Pollacco}, {Gillon}, {Queloz},
  {West}, {Bentley}, {Enoch}, {Haswell}, {Horne}, {Mayor}, {Pepe}, {Segransan},
  {Skillen}, {Udry}, \& {Wheatley}}]{hebb2010}
{Hebb}, L., {Collier-Cameron}, A., {Triaud}, A.~H.~M.~J., {et~al.} 2010, \apj,
  708, 224, \dodoi{10.1088/0004-637X/708/1/224}

\bibitem[{{H{\'e}brard} {et~al.}(2013){H{\'e}brard}, {Collier Cameron},
  {Brown}, {D{\'\i}az}, {Faedi}, {Smalley}, {Anderson}, {Armstrong}, {Barros},
  {Bento}, {Bouchy}, {Doyle}, {Enoch}, {G{\'o}mez Maqueo Chew}, {H{\'e}brard},
  {Hellier}, {Lendl}, {Lister}, {Maxted}, {McCormac}, {Moutou}, {Pollacco},
  {Queloz}, {Santerne}, {Skillen}, {Southworth}, {Tregloan-Reed}, {Triaud},
  {Udry}, {Vanhuysse}, {Watson}, {West}, \& {Wheatley}}]{hebrard2013}
{H{\'e}brard}, G., {Collier Cameron}, A., {Brown}, D.~J.~A., {et~al.} 2013,
  \aap, 549, A134, \dodoi{10.1051/0004-6361/201220363}

\bibitem[{{Heller}(2019)}]{heller2019}
{Heller}, R. 2019, \aap, 628, A42, \dodoi{10.1051/0004-6361/201833486}

\bibitem[{{Hellier} {et~al.}(2012){Hellier}, {Anderson}, {Collier Cameron},
  {Doyle}, {Fumel}, {Gillon}, {Jehin}, {Lendl}, {Maxted}, {Pepe}, {Pollacco},
  {Queloz}, {S{\'e}gransan}, {Smalley}, {Smith}, {Southworth}, {Triaud},
  {Udry}, \& {West}}]{hellier2012}
{Hellier}, C., {Anderson}, D.~R., {Collier Cameron}, A., {et~al.} 2012, \mnras,
  426, 739, \dodoi{10.1111/j.1365-2966.2012.21780.x}

\bibitem[{{Hellier} {et~al.}(2014){Hellier}, {Anderson}, {Collier Cameron},
  {Delrez}, {Gillon}, {Jehin}, {Lendl}, {Maxted}, {Pepe}, {Pollacco}, {Queloz},
  {S{\'e}gransan}, {Smalley}, {Smith}, {Southworth}, {Triaud}, {Udry}, \&
  {West}}]{hellier2014}
---. 2014, \mnras, 440, 1982, \dodoi{10.1093/mnras/stu410}

\bibitem[{{Hellier} {et~al.}(2017){Hellier}, {Anderson}, {Collier Cameron},
  {Delrez}, {Gillon}, {Jehin}, {Lendl}, {Maxted}, {Neveu-VanMalle}, {Pepe},
  {Pollacco}, {Queloz}, {S{\'e}gransan}, {Smalley}, {Southworth}, {Triaud},
  {Udry}, {Wagg}, \& {West}}]{hellier2017}
---. 2017, \mnras, 465, 3693, \dodoi{10.1093/mnras/stw3005}

\bibitem[{{Hellier} {et~al.}(2019){Hellier}, {Anderson}, {Bouchy}, {Burdanov},
  {Collier Cameron}, {Delrez}, {Gillon}, {Jehin}, {Lendl}, {Nielsen}, {Maxted},
  {Pepe}, {Pollacco}, {Queloz}, {S{\'e}gransan}, {Smalley}, {Triaud}, {Udry},
  \& {West}}]{hellier2019}
{Hellier}, C., {Anderson}, D.~R., {Bouchy}, F., {et~al.} 2019, \mnras, 482,
  1379, \dodoi{10.1093/mnras/sty2741}

\bibitem[{{Irwin} {et~al.}(2007){Irwin}, {Hodgkin}, {Aigrain}, {Hebb},
  {Bouvier}, {Clarke}, {Moraux}, \& {Bramich}}]{irwin2007}
{Irwin}, J., {Hodgkin}, S., {Aigrain}, S., {et~al.} 2007, \mnras, 377, 741,
  \dodoi{10.1111/j.1365-2966.2007.11640.x}

\bibitem[{{Jackson} {et~al.}(2009){Jackson}, {Barnes}, \&
  {Greenberg}}]{jackson2009}
{Jackson}, B., {Barnes}, R., \& {Greenberg}, R. 2009, \apj, 698, 1357,
  \dodoi{10.1088/0004-637X/698/2/1357}

\bibitem[{{Johns} {et~al.}(2019){Johns}, {Reed}, {Rodriguez}, {Pepper},
  {Stassun}, {Penev}, {Gaudi}, {Labadie-Bartz}, {Fulton}, {Quinn}, {Eastman},
  {Ciardi}, {Hirsch}, {Stevens}, {Stevens}, {Oberst}, {Cohen}, {Jensen},
  {Benni}, {Villanueva}, {Murawski}, {Bieryla}, {Latham}, {Vanaverbeke},
  {Dubois}, {Rau}, {Logie}, {Rauenzahn}, {Wittenmyer}, {Zambelli}, {Bayliss},
  {Beatty}, {Collins}, {Col{\'o}n}, {Curtis}, {Evans}, {Gregorio}, {James},
  {Depoy}, {Johnson}, {Joner}, {Kasper}, {Khakpash}, {Kielkopf}, {Kuhn},
  {Lund}, {Manner}, {Marshall}, {McLeod}, {Penny}, {Relles}, {Siverd},
  {Stephens}, {Stockdale}, {Tan}, {Trueblood}, {Trueblood}, \&
  {Yao}}]{johns2019}
{Johns}, D., {Reed}, P.~A., {Rodriguez}, J.~E., {et~al.} 2019, \aj, 158, 78,
  \dodoi{10.3847/1538-3881/ab24c7}

\bibitem[{{Kawaler}(1988)}]{kawaler1988}
{Kawaler}, S.~D. 1988, \apj, 333, 236, \dodoi{10.1086/166740}

\bibitem[{{Kley} \& {Nelson}(2012)}]{kley2012}
{Kley}, W., \& {Nelson}, R.~P. 2012, \araa, 50, 211,
  \dodoi{10.1146/annurev-astro-081811-125523}

\bibitem[{{Lai}(2012)}]{lai2012}
{Lai}, D. 2012, \mnras, 423, 486, \dodoi{10.1111/j.1365-2966.2012.20893.x}

\bibitem[{{Lai} {et~al.}(2011){Lai}, {Foucart}, \& {Lin}}]{lai2011}
{Lai}, D., {Foucart}, F., \& {Lin}, D. N.~C. 2011, \mnras, 412, 2790,
  \dodoi{10.1111/j.1365-2966.2010.18127.x}

\bibitem[{{Lanza} {et~al.}(2009){Lanza}, {Pagano}, {Leto}, {Messina},
  {Aigrain}, {Alonso}, {Auvergne}, {Baglin}, {Barge}, {Bonomo}, {Boumier},
  {Collier Cameron}, {Comparato}, {Cutispoto}, {de Medeiros}, {Foing},
  {Kaiser}, {Moutou}, {Parihar}, {Silva-Valio}, \& {Weiss}}]{lanza2009}
{Lanza}, A.~F., {Pagano}, I., {Leto}, G., {et~al.} 2009, \aap, 493, 193,
  \dodoi{10.1051/0004-6361:200810591}

\bibitem[{{Lendl} {et~al.}(2019){Lendl}, {Anderson}, {Bonfanti}, {Bouchy},
  {Burdanov}, {Collier Cameron}, {Delrez}, {Gillon}, {Hellier}, {Jehin},
  {Maxted}, {Nielsen}, {Pepe}, {Pollacco}, {Queloz}, {S{\'e}gransan},
  {Southworth}, {Smalley}, {Thompson}, {Turner}, {Triaud}, {Udry}, \&
  {West}}]{lendl2019}
{Lendl}, M., {Anderson}, D.~R., {Bonfanti}, A., {et~al.} 2019, \mnras, 482,
  301, \dodoi{10.1093/mnras/sty2667}

\bibitem[{{Li} \& {Winn}(2016)}]{li2016}
{Li}, G., \& {Winn}, J.~N. 2016, \apj, 818, 5,
  \dodoi{10.3847/0004-637X/818/1/5}

\bibitem[{{Lin} {et~al.}(1996){Lin}, {Bodenheimer}, \& {Richardson}}]{lin1996}
{Lin}, D.~N.~C., {Bodenheimer}, P., \& {Richardson}, D.~C. 1996, \nat, 380,
  606, \dodoi{10.1038/380606a0}

\bibitem[{{Lin} \& {Ogilvie}(2017)}]{lin2017}
{Lin}, Y., \& {Ogilvie}, G.~I. 2017, \mnras, 468, 1387,
  \dodoi{10.1093/mnras/stx540}

\bibitem[{{Lister} {et~al.}(2009){Lister}, {Anderson}, {Gillon}, {Hebb},
  {Smalley}, {Triaud}, {Collier Cameron}, {Wilson}, {West}, {Bentley},
  {Christian}, {Enoch}, {Haswell}, {Hellier}, {Horne}, {Irwin}, {Joshi},
  {Kane}, {Mayor}, {Maxted}, {Norton}, {Parley}, {Pepe}, {Pollacco}, {Queloz},
  {Ryans}, {Segransan}, {Skillen}, {Street}, {Todd}, {Udry}, \&
  {Wheatley}}]{lister2009}
{Lister}, T.~A., {Anderson}, D.~R., {Gillon}, M., {et~al.} 2009, \apj, 703,
  752, \dodoi{10.1088/0004-637X/703/1/752}

\bibitem[{{Maciejewski} {et~al.}(2016){Maciejewski}, {Dimitrov},
  {Fern{\'a}ndez}, {Sota}, {Nowak}, {Ohlert}, {Nikolov}, {Bukowiecki}, {Hinse},
  {Pall{\'e}}, {Tingley}, {Kjurkchieva}, {Lee}, \& {Lee}}]{Maciejewski+2016}
{Maciejewski}, G., {Dimitrov}, D., {Fern{\'a}ndez}, M., {et~al.} 2016, \aap,
  588, L6, \dodoi{10.1051/0004-6361/201628312}

\bibitem[{{Mancini} {et~al.}(2015){Mancini}, {Esposito}, {Covino}, {Raia},
  {Southworth}, {Tregloan-Reed}, {Biazzo}, {Bonomo}, {Desidera}, {Lanza},
  {Maciejewski}, {Poretti}, {Sozzetti}, {Borsa}, {Bruni}, {Ciceri}, {Claudi},
  {Cosentino}, {Gratton}, {Martinez Fiorenzano}, {Lodato}, {Lorenzi},
  {Marzari}, {Murabito}, {Affer}, {Bignamini}, {Bedin}, {Boccato}, {Damasso},
  {Henning}, {Maggio}, {Micela}, {Molinari}, {Pagano}, {Piotto}, {Rainer},
  {Scandariato}, {Smareglia}, \& {Zanmar Sanchez}}]{mancini2015}
{Mancini}, L., {Esposito}, M., {Covino}, E., {et~al.} 2015, \aap, 579, A136,
  \dodoi{10.1051/0004-6361/201526030}

\bibitem[{{Mancini} {et~al.}(2018){Mancini}, {Esposito}, {Covino},
  {Southworth}, {Biazzo}, {Bruni}, {Ciceri}, {Evans}, {Lanza}, {Poretti},
  {Sarkis}, {Smith}, {Brogi}, {Affer}, {Benatti}, {Bignamini}, {Boccato},
  {Bonomo}, {Borsa}, {Carleo}, {Claudi}, {Cosentino}, {Damasso}, {Desidera},
  {Giacobbe}, {Gonz{\'a}lez-{\'A}lvarez}, {Gratton}, {Harutyunyan}, {Leto},
  {Maggio}, {Malavolta}, {Maldonado}, {Martinez-Fiorenzano}, {Masiero},
  {Micela}, {Molinari}, {Nascimbeni}, {Pagano}, {Pedani}, {Piotto}, {Rainer},
  {Scandariato}, {Smareglia}, {Sozzetti}, {Andreuzzi}, \&
  {Henning}}]{mancini2018}
---. 2018, \aap, 613, A41, \dodoi{10.1051/0004-6361/201732234}

\bibitem[{{Mathis}(2015)}]{mathis2015}
{Mathis}, S. 2015, \aap, 580, L3, \dodoi{10.1051/0004-6361/201526472}

\bibitem[{{Matsumura} {et~al.}(2010){Matsumura}, {Peale}, \&
  {Rasio}}]{matsumura2010}
{Matsumura}, S., {Peale}, S.~J., \& {Rasio}, F.~A. 2010, \apj, 725, 1995,
  \dodoi{10.1088/0004-637X/725/2/1995}

\bibitem[{{Maxted} {et~al.}(2015){Maxted}, {Serenelli}, \&
  {Southworth}}]{maxted2015}
{Maxted}, P.~F.~L., {Serenelli}, A.~M., \& {Southworth}, J. 2015, \aap, 577,
  A90, \dodoi{10.1051/0004-6361/201525774}

\bibitem[{{Maxted} {et~al.}(2011){Maxted}, {Anderson}, {Collier Cameron},
  {Hellier}, {Queloz}, {Smalley}, {Street}, {Triaud}, {West}, {Gillon},
  {Lister}, {Pepe}, {Pollacco}, {S{\'e}gransan}, {Smith}, \&
  {Udry}}]{maxted2011}
{Maxted}, P.~F.~L., {Anderson}, D.~R., {Collier Cameron}, A., {et~al.} 2011,
  \pasp, 123, 547, \dodoi{10.1086/660007}

\bibitem[{{Mazeh} {et~al.}(2015){Mazeh}, {Perets}, {McQuillan}, \&
  {Goldstein}}]{mazeh2015}
{Mazeh}, T., {Perets}, H.~B., {McQuillan}, A., \& {Goldstein}, E.~S. 2015,
  \apj, 801, 3, \dodoi{10.1088/0004-637X/801/1/3}

\bibitem[{{Milliman} {et~al.}(2014){Milliman}, {Mathieu}, {Geller}, {Gosnell},
  {Meibom}, \& {Platais}}]{milliman2014}
{Milliman}, K.~E., {Mathieu}, R.~D., {Geller}, A.~M., {et~al.} 2014, \aj, 148,
  38, \dodoi{10.1088/0004-6256/148/2/38}

\bibitem[{{Morton}(2015)}]{morton2015}
{Morton}, T.~D. 2015, {isochrones: Stellar model grid package}.
\newblock \doeprint{1503.010}

\bibitem[{{Nielsen} {et~al.}(2019){Nielsen}, {Bouchy}, {Turner}, {Anderson},
  {Barkaoui}, {Benkhaldoun}, {Burdanov}, {Collier Cameron}, {Delrez}, {Gillon},
  {Ducrot}, {Hellier}, {Jehin}, {Lendl}, {Maxted}, {Pepe}, {Pollacco},
  {Pozuelos}, {Queloz}, {S{\'e}gransan}, {Smalley}, {Triaud}, {Udry}, \&
  {West}}]{nielsen2019}
{Nielsen}, L.~D., {Bouchy}, F., {Turner}, O.~D., {et~al.} 2019, \mnras, 489,
  2478, \dodoi{10.1093/mnras/stz2351}

\bibitem[{{O'Donovan} {et~al.}(2006){O'Donovan}, {Charbonneau}, {Mandushev},
  {Dunham}, {Latham}, {Torres}, {Sozzetti}, {Brown}, {Trauger}, {Belmonte},
  {Rabus}, {Almenara}, {Alonso}, {Deeg}, {Esquerdo}, {Falco}, {Hillenbrand},
  {Roussanova}, {Stefanik}, \& {Winn}}]{odonovan2006}
{O'Donovan}, F.~T., {Charbonneau}, D., {Mandushev}, G., {et~al.} 2006, \apjl,
  651, L61, \dodoi{10.1086/509123}

\bibitem[{{Ogilvie}(2013)}]{ogilvie2013}
{Ogilvie}, G.~I. 2013, \mnras, 429, 613, \dodoi{10.1093/mnras/sts362}

\bibitem[{{Ogilvie}(2014)}]{ogilvie2014}
---. 2014, \araa, 52, 171, \dodoi{10.1146/annurev-astro-081913-035941}

\bibitem[{{Ogilvie} \& {Lesur}(2012)}]{ogilvie2012}
{Ogilvie}, G.~I., \& {Lesur}, G. 2012, \mnras, 422, 1975,
  \dodoi{10.1111/j.1365-2966.2012.20630.x}

\bibitem[{{Ogilvie} \& {Lin}(2007)}]{ogilvie2007}
{Ogilvie}, G.~I., \& {Lin}, D.~N.~C. 2007, \apj, 661, 1180,
  \dodoi{10.1086/515435}

\bibitem[{{Penev} {et~al.}(2009){Penev}, {Barranco}, \& {Sasselov}}]{penev2009}
{Penev}, K., {Barranco}, J., \& {Sasselov}, D. 2009, \apj, 705, 285,
  \dodoi{10.1088/0004-637X/705/1/285}

\bibitem[{{Penev} {et~al.}(2018){Penev}, {Bouma}, {Winn}, \&
  {Hartman}}]{penev2018}
{Penev}, K., {Bouma}, L.~G., {Winn}, J.~N., \& {Hartman}, J.~D. 2018, \aj, 155,
  165, \dodoi{10.3847/1538-3881/aaaf71}

\bibitem[{{Penev} {et~al.}(2014){Penev}, {Zhang}, \& {Jackson}}]{penev2014}
{Penev}, K., {Zhang}, M., \& {Jackson}, B. 2014, \pasp, 126, 553,
  \dodoi{10.1086/677042}

\bibitem[{{Penev} {et~al.}(2016){Penev}, {Hartman}, {Bakos}, {Ciceri}, {Brahm},
  {Bayliss}, {Bento}, {Jord{\'a}n}, {Csubry}, {Bhatti}, {de Val-Borro},
  {Espinoza}, {Zhou}, {Mancini}, {Rabus}, {Suc}, {Henning}, {Schmidt}, {Noyes},
  {L{\'a}z{\'a}r}, {Papp}, \& {S{\'a}ri}}]{penev2016}
{Penev}, K., {Hartman}, J.~D., {Bakos}, G.~{\'A}., {et~al.} 2016, \aj, 152,
  127, \dodoi{10.3847/0004-6256/152/5/127}

\bibitem[{{Rogers} \& {Lin}(2013)}]{rogers2013}
{Rogers}, T.~M., \& {Lin}, D.~N.~C. 2013, \apjl, 769, L10,
  \dodoi{10.1088/2041-8205/769/1/L10}

\bibitem[{{Sanchis-Ojeda} {et~al.}(2011){Sanchis-Ojeda}, {Winn}, {Holman},
  {Carter}, {Osip}, \& {Fuentes}}]{sanchis-Ojeda2011}
{Sanchis-Ojeda}, R., {Winn}, J.~N., {Holman}, M.~J., {et~al.} 2011, \apj, 733,
  127, \dodoi{10.1088/0004-637X/733/2/127}

\bibitem[{{Sanchis-Ojeda} {et~al.}(2015){Sanchis-Ojeda}, {Winn}, {Dai},
  {Howard}, {Isaacson}, {Marcy}, {Petigura}, {Sinukoff}, {Weiss}, {Albrecht},
  {Hirano}, \& {Rogers}}]{sanchis-Ojeda2015}
{Sanchis-Ojeda}, R., {Winn}, J.~N., {Dai}, F., {et~al.} 2015, \apjl, 812, L11,
  \dodoi{10.1088/2041-8205/812/1/L11}

\bibitem[{{Santerne} {et~al.}(2011){Santerne}, {Bonomo}, {H{\'e}brard},
  {Deleuil}, {Moutou}, {Almenara}, {Bouchy}, \& {D{\'\i}az}}]{santerne2011}
{Santerne}, A., {Bonomo}, A.~S., {H{\'e}brard}, G., {et~al.} 2011, \aap, 536,
  A70, \dodoi{10.1051/0004-6361/201117807}

\bibitem[{{Santos} {et~al.}(2013){Santos}, {Sousa}, {Mortier}, {Neves},
  {Adibekyan}, {Tsantaki}, {Delgado Mena}, {Bonfils}, {Israelian}, {Mayor}, \&
  {Udry}}]{santos2013}
{Santos}, N.~C., {Sousa}, S.~G., {Mortier}, A., {et~al.} 2013, \aap, 556, A150,
  \dodoi{10.1051/0004-6361/201321286}

\bibitem[{{Schlaufman}(2010)}]{Schlaufman2010}
{Schlaufman}, K.~C. 2010, \apj, 719, 602, \dodoi{10.1088/0004-637X/719/1/602}

\bibitem[{{Simpson} {et~al.}(2011){Simpson}, {Faedi}, {Barros}, {Brown},
  {Collier Cameron}, {Hebb}, {Pollacco}, {Smalley}, {Todd}, {Butters},
  {H{\'e}brard}, {McCormac}, {Miller}, {Santerne}, {Street}, {Skillen},
  {Triaud}, {Anderson}, {Bento}, {Boisse}, {Bouchy}, {Enoch}, {Haswell},
  {Hellier}, {Holmes}, {Horne}, {Keenan}, {Lister}, {Maxted}, {Moulds},
  {Moutou}, {Norton}, {Parley}, {Pepe}, {Queloz}, {Segransan}, {Smith},
  {Stempels}, {Udry}, {Watson}, {West}, \& {Wheatley}}]{simpson2011}
{Simpson}, E.~K., {Faedi}, F., {Barros}, S.~C.~C., {et~al.} 2011, \aj, 141, 8,
  \dodoi{10.1088/0004-6256/141/1/8}

\bibitem[{{Skumanich}(1972)}]{skumanich1972}
{Skumanich}, A. 1972, \apj, 171, 565, \dodoi{10.1086/151310}

\bibitem[{{Smalley} {et~al.}(2011){Smalley}, {Anderson}, {Collier Cameron},
  {Hellier}, {Lendl}, {Maxted}, {Queloz}, {Triaud}, {West}, {Bentley}, {Enoch},
  {Gillon}, {Lister}, {Pepe}, {Pollacco}, {Segransan}, {Smith}, {Southworth},
  {Udry}, {Wheatley}, {Wood}, \& {Bento}}]{smalley2011}
{Smalley}, B., {Anderson}, D.~R., {Collier Cameron}, A., {et~al.} 2011, \aap,
  526, A130, \dodoi{10.1051/0004-6361/201015992}

\bibitem[{{Smith} {et~al.}(2012){Smith}, {Anderson}, {Collier Cameron},
  {Gillon}, {Hellier}, {Lendl}, {Maxted}, {Queloz}, {Smalley}, {Triaud},
  {West}, {Barros}, {Jehin}, {Pepe}, {Pollacco}, {Segransan}, {Southworth},
  {Street}, \& {Udry}}]{smith2012}
{Smith}, A.~M.~S., {Anderson}, D.~R., {Collier Cameron}, A., {et~al.} 2012,
  \aj, 143, 81, \dodoi{10.1088/0004-6256/143/4/81}

\bibitem[{{Southworth}(2011)}]{southworth2011}
{Southworth}, J. 2011, \mnras, 417, 2166,
  \dodoi{10.1111/j.1365-2966.2011.19399.x}

\bibitem[{{Southworth} {et~al.}(2016){Southworth}, {Tregloan-Reed}, {Andersen},
  {Calchi Novati}, {Ciceri}, {Colque}, {D'Ago}, {Dominik}, {Evans}, {Gu},
  {Herrera-Cordova}, {Hinse}, {J{\o}rgensen}, {Juncher}, {Kuffmeier},
  {Mancini}, {Peixinho}, {Popovas}, {Rabus}, {Skottfelt}, {Tronsgaard},
  {Unda-Sanzana}, {Wang}, {Wertz}, {Alsubai}, {Andersen}, {Bozza}, {Bramich},
  {Burgdorf}, {Damerdji}, {Diehl}, {Elyiv}, {Figuera Jaimes}, {Haugb{\o}lle},
  {Hundertmark}, {Kains}, {Kerins}, {Korhonen}, {Liebig}, {Mathiasen}, {Penny},
  {Rahvar}, {Scarpetta}, {Schmidt}, {Snodgrass}, {Starkey}, {Surdej}, {Vilela},
  {von Essen}, \& {Wang}}]{southworth2016}
{Southworth}, J., {Tregloan-Reed}, J., {Andersen}, M.~I., {et~al.} 2016,
  \mnras, 457, 4205, \dodoi{10.1093/mnras/stw279}

\bibitem[{{Stassun} \& {Torres}(2018)}]{stassun2018}
{Stassun}, K.~G., \& {Torres}, G. 2018, \apj, 862, 61,
  \dodoi{10.3847/1538-4357/aacafc}

\bibitem[{{Teitler} \& {K{\"o}nigl}(2014)}]{teitler2014}
{Teitler}, S., \& {K{\"o}nigl}, A. 2014, \apj, 786, 139,
  \dodoi{10.1088/0004-637X/786/2/139}

\bibitem[{{Teyssandier} {et~al.}(2019){Teyssandier}, {Lai}, \&
  {Vick}}]{teyssandier2019}
{Teyssandier}, J., {Lai}, D., \& {Vick}, M. 2019, \mnras, 486, 2265,
  \dodoi{10.1093/mnras/stz1011}

\bibitem[{{Tregloan-Reed} {et~al.}(2013){Tregloan-Reed}, {Southworth}, \&
  {Tappert}}]{tregloan2013}
{Tregloan-Reed}, J., {Southworth}, J., \& {Tappert}, C. 2013, \mnras, 428,
  3671, \dodoi{10.1093/mnras/sts306}

\bibitem[{{Triaud}(2018)}]{Triaud2018}
{Triaud}, A. H.~M.~J. 2018, {The Rossiter-McLaughlin Effect in Exoplanet
  Research}, ed. H.~J. {Deeg} \& J.~A. {Belmonte}, 2

\bibitem[{{Triaud} {et~al.}(2010){Triaud}, {Collier Cameron}, {Queloz},
  {Anderson}, {Gillon}, {Hebb}, {Hellier}, {Loeillet}, {Maxted}, {Mayor},
  {Pepe}, {Pollacco}, {S{\'e}gransan}, {Smalley}, {Udry}, {West}, \&
  {Wheatley}}]{triaud2010}
{Triaud}, A.~H.~M.~J., {Collier Cameron}, A., {Queloz}, D., {et~al.} 2010,
  \aap, 524, A25, \dodoi{10.1051/0004-6361/201014525}

\bibitem[{{Turner} {et~al.}(2016){Turner}, {Anderson}, {Collier Cameron},
  {Delrez}, {Evans}, {Gillon}, {Hellier}, {Jehin}, {Lendl}, {Maxted}, {Pepe},
  {Pollacco}, {Queloz}, {S{\'e}gransan}, {Smalley}, {Smith}, {Triaud}, {Udry},
  \& {West}}]{turner2016}
{Turner}, O.~D., {Anderson}, D.~R., {Collier Cameron}, A., {et~al.} 2016,
  \pasp, 128, 064401, \dodoi{10.1088/1538-3873/128/964/064401}

\bibitem[{{Valsecchi} \& {Rasio}(2014)}]{Valsecchi+2014}
{Valsecchi}, F., \& {Rasio}, F.~A. 2014, \apj, 786, 102,
  \dodoi{10.1088/0004-637X/786/2/102}

\bibitem[{{Vick} {et~al.}(2019){Vick}, {Lai}, \& {Anderson}}]{vick2019}
{Vick}, M., {Lai}, D., \& {Anderson}, K.~R. 2019, \mnras, 484, 5645,
  \dodoi{10.1093/mnras/stz354}

\bibitem[{{Weinberg} {et~al.}(2017){Weinberg}, {Sun}, {Arras}, \&
  {Essick}}]{Weinberg+2017}
{Weinberg}, N.~N., {Sun}, M., {Arras}, P., \& {Essick}, R. 2017, \apjl, 849,
  L11, \dodoi{10.3847/2041-8213/aa9113}

\bibitem[{{Wilson} {et~al.}(2006){Wilson}, {Enoch}, {Christian}, {Clarkson},
  {Collier Cameron}, {Deeg}, {Evans}, {Haswell}, {Hellier}, {Hodgkin}, {Horne},
  {Irwin}, {Kane}, {Lister}, {Maxted}, {Norton}, {Pollacco}, {Skillen},
  {Street}, {West}, \& {Wheatley}}]{wilson2006}
{Wilson}, D.~M., {Enoch}, B., {Christian}, D.~J., {et~al.} 2006, \pasp, 118,
  1245, \dodoi{10.1086/507957}

\bibitem[{{Wilson} {et~al.}(2008){Wilson}, {Gillon}, {Hellier}, {Maxted},
  {Pepe}, {Queloz}, {Anderson}, {Collier Cameron}, {Smalley}, {Lister},
  {Bentley}, {Blecha}, {Christian}, {Enoch}, {Haswell}, {Hebb}, {Horne},
  {Irwin}, {Joshi}, {Kane}, {Marmier}, {Mayor}, {Parley}, {Pollacco}, {Pont},
  {Ryans}, {Segransan}, {Skillen}, {Street}, {Udry}, {West}, \&
  {Wheatley}}]{wilson2008}
{Wilson}, D.~M., {Gillon}, M., {Hellier}, C., {et~al.} 2008, \apjl, 675, L113,
  \dodoi{10.1086/586735}

\bibitem[{{Winn} {et~al.}(2010){Winn}, {Fabrycky}, {Albrecht}, \&
  {Johnson}}]{winn2010}
{Winn}, J.~N., {Fabrycky}, D., {Albrecht}, S., \& {Johnson}, J.~A. 2010, \apjl,
  718, L145, \dodoi{10.1088/2041-8205/718/2/L145}

\bibitem[{{Winn} \& {Fabrycky}(2015)}]{WinnFabrycky2015}
{Winn}, J.~N., \& {Fabrycky}, D.~C. 2015, \araa, 53, 409,
  \dodoi{10.1146/annurev-astro-082214-122246}

\bibitem[{{Winn} {et~al.}(2008){Winn}, {Johnson}, {Narita}, {Suto}, {Turner},
  {Fischer}, {Butler}, {Vogt}, {O'Donovan}, \& {Gaudi}}]{winn2008}
{Winn}, J.~N., {Johnson}, J.~A., {Narita}, N., {et~al.} 2008, \apj, 682, 1283,
  \dodoi{10.1086/589235}

\bibitem[{{Wu} \& {Lithwick}(2011)}]{wu2011}
{Wu}, Y., \& {Lithwick}, Y. 2011, \apj, 735, 109,
  \dodoi{10.1088/0004-637X/735/2/109}

\bibitem[{{Wu} \& {Murray}(2003)}]{wu2003}
{Wu}, Y., \& {Murray}, N. 2003, \apj, 589, 605, \dodoi{10.1086/374598}

\bibitem[{{Xue} {et~al.}(2014){Xue}, {Suto}, {Taruya}, {Hirano}, {Fujii}, \&
  {Masuda}}]{xue2014}
{Xue}, Y., {Suto}, Y., {Taruya}, A., {et~al.} 2014, \apj, 784, 66,
  \dodoi{10.1088/0004-637X/784/1/66}

\bibitem[{{Yee} {et~al.}(2020){Yee}, {Winn}, {Knutson}, {Patra},
  {Vissapragada}, {Zhang}, {Holman}, {Shporer}, \& {Wright}}]{yee2020}
{Yee}, S.~W., {Winn}, J.~N., {Knutson}, H.~A., {et~al.} 2020, \apjl, 888, L5,
  \dodoi{10.3847/2041-8213/ab5c16}

\bibitem[{{Zahn}(1966)}]{zahn1966}
{Zahn}, J.~P. 1966, Annales d'Astrophysique, 29, 313

\bibitem[{{Zahn}(1977)}]{zahn1977}
---. 1977, \aap, 500, 121

\end{thebibliography}

\appendix
\restartappendixnumbering

\section{Stellar Age Calculations}
\label{sec:ages}
We use the \texttt{Isochrones} package to calculate stellar ages, including as input the stellar effective temperature and metallicity from SWEET-Cat, the stellar density from the transit light curve parameters (see equation \ref{eq:density}), the Gaia parallax and photometric bands ($G, RP, BP$), 2MASS and WISE 123 photometric bands (as listed for each system on the NASA Exoplanet Archive), and the visual extinction at the sky coordinates and distance, using the 3D dust map MWDUST \citep{bovy2016}.

We adopt the uncertainties from the literature for all the input parameters, with the following modifications. We impose minimum uncertainties of $100$\,K and 0.1\,dex in the stellar effective temperatures and metallicities, due to the uncertainties in stellar atmospheric model parameters, and a minimum $10 \%$ uncertainty in the stellar density. To account for systematic effects, we impose a minimum uncertainty of 0.1 mas in the Gaia parallaxes (\citealt{gaia2018}; see also \citealt{stassun2018}). On all photometry measurements, we impose minimum errors of $0.01$ mag. In practice, this choice affects only the Gaia photometric uncertainties, and is consistent with expected Gaia-band systematic errors \citep{evans2018}. For a given system, we exclude from our input parameters any photometric bands having an uncertainty larger than $0.1$ mag, or
for which the value or uncertainty is missing from the NASA Exoplanet Archive. The choices for these errors reflect our aim to obtain realistic and uniformly-derived age posteriors.

The final sample consists of 46 hot Jupiter systems, and is given in Table \ref{table}. Note that we have discarded CoRoT-26 because it has a negative Gaia parallax, and K2-238 because of a large parallax error, leading to poor age constraints. For some systems (CoRoT-2, CoRoT-13, CoRoT-17, KELT-8, WASP-114, WASP-123, WASP-135, WASP-70 A, and WASP-77 A), including the 2MASS and WISE photometry as input for the \texttt{Isochrones} package results in posterior distributions extremely inconsistent with the input parameters and uncertainties. This may result from binary companions or contamination from background stars. For these systems, we repeated the age calculations including only the Gaia photometry, and we retained a given system in our sample if the new posteriors adequately encompass the input parameters. We remove WASP-135 from our sample, because \texttt{Isochrones} returns an implausibly low age and large mass ($t_{\rm age} \lesssim 10$ Myr and $M_\star \sim 3 \msun$) regardless of the input photometry. We similarly remove K2-30. We remove WASP-77 A from the sample, because although removing the 2MASS and WISE photometry yielded spectroscopic parameter posteriors consistent with the observed values, the resulting age posterior was extremely broad and uninformative. For Kepler-41, \texttt{Isochrones} predicts a much larger metallicity ($\feh \sim 0.4$ dex) compared to the SWEET-CAT value $\feh = -0.09 \pm 0.16$. Since \cite{bonomo2015} found $\feh = 0.38 \pm 0.11$ for Kepler-41, we opted to use the spectroscopic parameters from \cite{bonomo2015} instead of those from SWEET-CAT for this particular system.

In Figure A1, the values of $\teff$, $\feh$ and $\rho_\star$ obtained from the posterior distributions returned by \texttt{Isochrones} are plotted against the input values. In general, the input parameters are well-matched. For six systems (CoRoT-13,WASP-57, WASP-60, K2-30, WASP-16, and WASP-34), the \texttt{Isochrones} density is substantially lower than the input density, estimated from equation (\ref{eq:density}). Equation (\ref{eq:density}) may predict an inaccurate density if the orbit is eccentric or if the tabulated transit impact parameter is incorrect.

\begin{figure*}
\centering 
\includegraphics[width=0.9\textwidth]{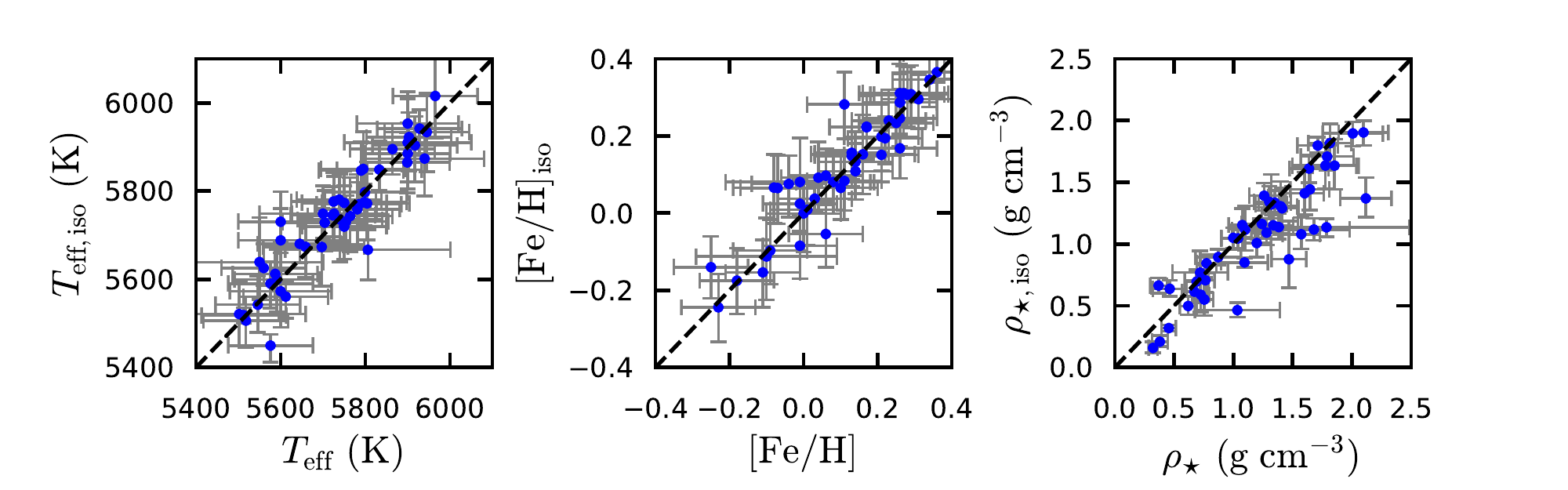}
\caption{$\teff$, $\feh$, and $\rho_\star$ output by \texttt{Isochrones}, compared to the corresponding input values (as obtained from the SWEET-CAT catalogue and equation [\ref{eq:density}]). The y-axis error bars and values indicate the 16th, 50th, and 84th percentiles of the posterior distributions obtained from the \texttt{Isochrones}, while the x-axis error bars and values indicate our input values and adopted uncertainties (see text). In general, the input values fit comfortably within the posterior distributions. For six systems, the densities obtained by \texttt{Isochrones} are significantly lower than the input values, possibly indicating an eccentric orbit or something amiss with the transit impact parameter.}
\label{fig:specparams}
\end{figure*}

To investigate this issue, we have repeated the \texttt{Isochrones} calculation, using the $\log g$ reported in SWEET-Cat in place of the stellar density. Figure A2 shows the results. For CoRoT-13,K2-30,WASP-16, and WASP-34, the $\log g$ posteriors more closely reflect the input $\log g$, while the $\log g$ posteriors for WASP-57 and WASP-60 remain underestimated. This may indicate a modest eccentricity in the former systems.

\begin{figure}
\centering 
\includegraphics[scale=0.5]{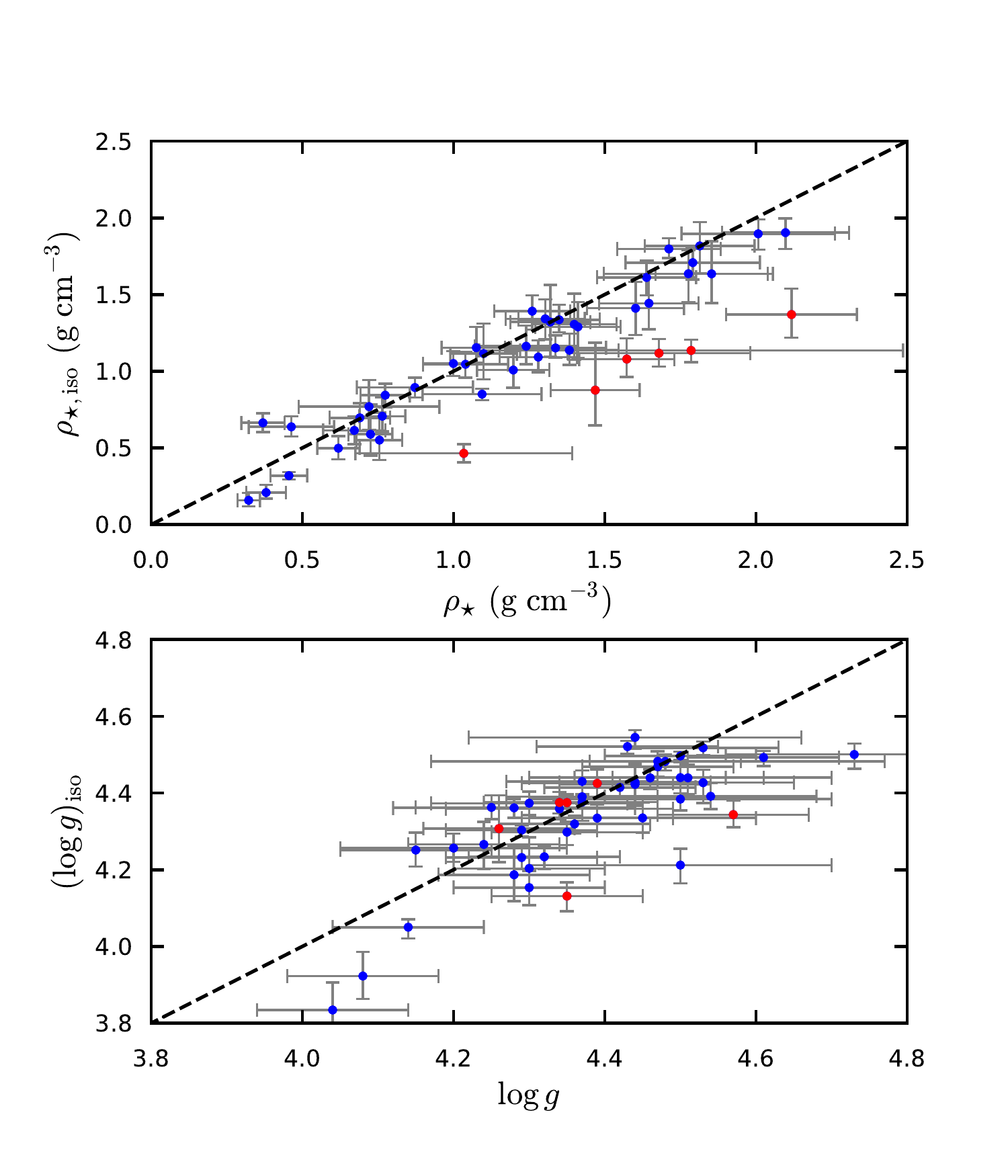}
\caption{{\it Top panel:} Comparison of the density obtained from \texttt{Isochrones} to the input density determined from equation (\ref{eq:density}), as depicted in the rightmost panel of Fig.~\ref{fig:specparams}. Highlighted in red are the six systems with underpredicted densities returned by \texttt{Isochrones}. {\it Bottom panel:} Results of a separate \texttt{Isochrones} calculation, using $\log g$ instead of $\rho_\star$. The six systems with underpredicted stellar densities from the top panel are shown in red. Only two of these six systems (WASP-57 and WASP-60) exhibit a correspondingly low $\log g$, although there is a slight tendency for \texttt{Isochrones} to predict lower $\log g$.}
\label{fig:rho_logg}
\end{figure}

As a consistency check, we compare the ages obtained from using $\rho_\star$ with the ages from using $\log g$. The ages determined using the two different methods are in good agreement. Our adopted minimum uncertainties for $\rho_\star$ (10 \%) and $\log \, g$ (0.1 dex) yield similarly broad age posteriors.

\clearpage

\startlongtable
\begin{deluxetable*}{lLLLLLLLl}
\tablecaption{Final sample of hot Jupiters orbiting cool host stars, and the stellar ages derived in this paper. \label{table}}
\tabletypesize{\small}
\startdata
\tablehead{\colhead{Name} & \colhead{$\mplanet$ ($\mjup$)} & \colhead{$P_{\rm orb}$ (d)} & \colhead{$\teff$ (K)}  & \colhead{$P_{\star}^{(\rm phot)}$ (d)} & \colhead{$P_\star^{(v \sin i)}$ (d)} & \colhead{$\lambda$ ($^\circ$)} & \colhead{age (Gyr)} & \colhead{Refs.}} 
CoRoT-13 & 1.31 & 4.0 & 5945 & \nodata & 12.8 \pm 3.2 & \nodata & 5.6 ^{+1.9}_{-2.4}  & 1\\
CoRoT-17 & 2.43 & 3.8 & 5740 & \nodata & 17.9 \pm 2.1 & \nodata & 4.9 ^{+1.3}_{-1.1}  & 2\\
CoRoT-2 & 3.47 & 1.7 & 5697 & 4.5$\pm$0.02 & 4.2 \pm 0.2 & 7.2 $\pm$ 4.5 & 3.7 ^{+4.0}_{-2.0}  & 3,4,5\\
HAT-P-13 & 0.85 & 2.9 & 5797 & \nodata & 47.5 \pm 10.9 & 1.9 $\pm$ 8.6 & 4.0 ^{+1.5}_{-0.4}  & 6,7\\
HAT-P-36 & 1.85 & 1.3 & 5560 & 15.3$\pm$0.4 & 14.7 \pm 4.4 & -14.0 $\pm$ 18.0 & 4.4 ^{+1.9}_{-1.8}  & 8,9\\
HAT-P-42 & 1.04 & 4.6 & 5903 & \nodata & 22.1 \pm 3.8 & \nodata & 3.9 ^{+0.9}_{-0.7}  & 10\\
HAT-P-43 & 0.66 & 3.3 & 5645 & \nodata & 23.2 \pm 4.9 & \nodata & 5.6 ^{+1.9}_{-2.2}  & 10\\
HAT-P-5 & 0.98 & 2.8 & 5833 & \nodata & 21.8 \pm 12.7 & \nodata & 4.2 ^{+1.6}_{-1.9}  & 11\\
HATS-16 & 3.27 & 2.7 & 5738 & 12.4$\pm$0.02 & 10.2 \pm 1.1 & \nodata & 7.7 ^{+1.4}_{-1.6}  & 12\\
HATS-18 & 1.98 & 0.8 & 5600 & 9.8$\pm$0.4 & 8.3 \pm 0.8 & \nodata & 4.8 ^{+3.0}_{-2.9}  & 13\\
HATS-57 & 3.15 & 2.4 & 5587 & 12.7$\pm$0.04 & 11.9 \pm 1.4 & \nodata & 1.7 ^{+1.6}_{-1.1}  & 14\\
KELT-23 A & 0.94 & 2.3 & 5899 & \nodata & 20.9 \pm 4.3 & \nodata & 6.0 ^{+2.1}_{-2.4}  & 15\\
KELT-8 & 0.66 & 3.2 & 5804 & \nodata & 20.0 \pm 8.2 & \nodata & 6.4 ^{+1.6}_{-1.4}  & 16\\
Kepler-17 & 2.45 & 1.5 & 5781 & 11.9$\pm$1.1 & 8.9 \pm 3.0 & 0.0 $\pm$ 15.0 & 2.2 ^{+2.3}_{-1.5}  & 17\\
Kepler-41 & 0.56 & 1.9 & 5660 & 20.9$\pm$0.31 & 10.9 \pm 3.6 & \nodata & 5.3 ^{+1.7}_{-1.3}  & 18,19\\
Kepler-412 & 0.94 & 1.7 & 5750 & 17.2$\pm$1.6 & 13.0 \pm 2.6 & \nodata & 5.2 ^{+1.4}_{-1.0}  & 20,21\\
Kepler-423 & 0.6 & 2.7 & 5790 & 22.0$\pm$0.15 & 19.2 \pm 3.9 & \nodata & 2.7 ^{+2.9}_{-1.9}  & 22,19\\
Kepler-6 & 0.67 & 3.2 & 5659 & 59.9$\pm$6.47 & 23.4 \pm 7.8 & \nodata & 7.1 ^{+1.4}_{-1.2}  & 23,19\\
TrES-2 & 1.49 & 2.5 & 5795 & 70.6$\pm$16.47 & 28.3 \pm 21.3 & -9.0 $\pm$ 12.0 & 6.1 ^{+2.3}_{-2.0}  & 24,19,25\\
WASP-114 & 1.77 & 1.5 & 5940 & \nodata & 11.3 \pm 1.3 & \nodata & 6.0 ^{+1.6}_{-1.3}  & 26\\
WASP-123 & 0.9 & 3.0 & 5764 & \nodata & 64.7 \pm 45.4 & \nodata & 6.7 ^{+2.1}_{-1.6}  & 27\\
WASP-141 & 2.69 & 3.3 & 5900 & \nodata & 17.8 \pm 3.8 & \nodata & 3.6 ^{+0.7}_{-0.9}  & 28\\
WASP-16 & 1.24 & 3.1 & 5726 & \nodata & 19.2 \pm 6.6 & -4.2 $^{+13.9}_{-11.0}$  & 5.8 ^{+2.2}_{-1.8}  & 29,30\\
WASP-164 & 2.13 & 1.8 & 5806 & 17.8$\pm$0.03 & \nodata & \nodata & 6.9 ^{+2.4}_{-1.5}  & 31\\
WASP-171 & 1.08 & 3.8 & 5965 & \nodata & 13.2 \pm 2.0 & \nodata & 4.1 ^{+0.7}_{-1.0}  & 32\\
WASP-173 A & 3.69 & 1.4 & 5700 & 7.9$\pm$0.1 & 9.2 \pm 0.6 & \nodata & 5.8 ^{+1.9}_{-1.4}  & 33\\
WASP-19 & 1.07 & 0.8 & 5591 & 11.8$\pm$0.09 & 10.9 \pm 0.7 & 1.0 $\pm$ 1.2 & 2.9 ^{+2.6}_{-1.7}  & 34,35\\
WASP-192 & 2.3 & 2.9 & 5900 & \nodata & 21.5 \pm 7.7 & \nodata & 5.5 ^{+1.8}_{-1.2}  & 33\\
WASP-34 & 0.56 & 4.3 & 5704 & \nodata & 32.9 \pm 14.4 & \nodata & 6.3 ^{+2.1}_{-2.1}  & 36\\
WASP-36 & 2.36 & 1.5 & 5928 & \nodata & 15.0 \pm 5.5 & \nodata & 2.4 ^{+2.6}_{-1.6}  & 37\\
WASP-37 & 1.8 & 3.6 & 5917 & \nodata & 21.1 \pm 14.1 & \nodata & 8.5 ^{+2.2}_{-2.0}  & 38\\
WASP-4 & 1.19 & 1.3 & 5513 & 22.5$\pm$3.3 & 20.5 \pm 9.3 & -1.0 $^{+14.0}_{-12.0}$  & 4.7 ^{+2.7}_{-1.7}  & 39,40,41\\
WASP-41 & 0.85 & 3.1 & 5546 & 18.6$\pm$1.5 & 15.9 \pm 1.7 & 6.0 $\pm$ 11.0 & 2.4 ^{+2.6}_{-1.6}  & 42,43,43,\\
WASP-44 & 0.87 & 2.4 & 5612 & \nodata & 14.4 \pm 4.4 & \nodata & 3.0 ^{+2.8}_{-1.9}  & 44\\
WASP-46 & 1.91 & 1.4 & 5725 & 16.0$\pm$1.0 & 22.9 \pm 14.5 & \nodata & 5.1 ^{+3.3}_{-2.9}  & 44\\
WASP-47 & 1.14 & 4.2 & 5576 & \nodata & 32.0 \pm 4.3 & 0.0 $\pm$ 24.0 & 6.7 ^{+1.6}_{-1.3}  & 45,46\\
WASP-5 & 1.58 & 1.6 & 5785 & \nodata & 17.1 \pm 5.3 & 12.0 $^{+8.0}_{-10.0}$  & 4.6 ^{+1.5}_{-1.7}  & 44,47\\
WASP-50 & 1.47 & 2.0 & 5518 & 16.3$\pm$0.5 & 16.3 \pm 4.9 & \nodata & 2.8 ^{+2.9}_{-2.4}  & 48\\
WASP-57 & 0.64 & 2.8 & 5600 & \nodata & 12.7 \pm 4.5 & \nodata & 11.0 ^{+1.6}_{-2.3}  & 49\\
WASP-60 & 0.55 & 4.3 & 5900 & \nodata & 17.4 \pm 4.7 & -129.0 $\pm$ 17.0 & 6.7 ^{+1.2}_{-1.5}  & 50,51\\
WASP-64 & 1.27 & 1.6 & 5550 & \nodata & 15.8 \pm 3.7 & \nodata & 9.4 ^{+2.3}_{-2.8}  & 52\\
WASP-65 & 1.55 & 2.3 & 5600 & \nodata & 14.2 \pm 2.1 & \nodata & 7.1 ^{+1.8}_{-1.9}  & 53\\
WASP-70 A & 0.59 & 3.7 & 5864 & \nodata & 34.3 \pm 8.0 & \nodata & 5.1 ^{+1.6}_{-1.3}  & 54\\
WASP-95 & 1.44 & 2.2 & 5799 & 20.7$\pm$2.7 & 20.1 \pm 4.2 & \nodata & 5.5 ^{+1.2}_{-1.0}  & 55\\
WASP-97 & 1.36 & 2.1 & 5723 & \nodata & 49.7 \pm 22.7 & \nodata & 2.1 ^{+3.4}_{-0.5}  & 55\\
XO-1 & 0.83 & 3.9 & 5754 & \nodata & 40.1 \pm 3.3 & \nodata & 2.4 ^{+2.0}_{-1.5}  & 56\\
\enddata
\tablerefs{(1) \cite{cabrera2010}, (2) \cite{csizmadia2011}, (3) \cite{alonso2008}, (4) \cite{lanza2009}, (5) \cite{bouchy2008}, (6) \cite{bakos2009}, (7) \cite{winn2010}, (8) \cite{bakos2012}, (9) \cite{mancini2015}, (10) \cite{boisse2013}, (11) \cite{bakos2007}, (12) \cite{ciceri2016}, (13) \cite{penev2016}, (14) \cite{espinoza2019}, (15) \cite{johns2019}, (16) \cite{fulton2015}, (17) \cite{desert2011}, (18) \cite{santerne2011}, (19) \cite{mazeh2015}, (20) \cite{borucki2011}, (21) \cite{deleuil2014}, (22) \cite{endl2014}, (23) \cite{dunham2010}, (24) \cite{odonovan2006}, (25) \cite{winn2008}, (26) \cite{barros2016}, (27) \cite{turner2016}, (28) \cite{hellier2017}, (29) \cite{lister2009}, (30) \cite{brown2012}, (31) \cite{lendl2019}, (32) \cite{nielsen2019}, (33) \cite{hellier2019}, (34) \cite{hebb2010}, (35) \cite{tregloan2013}, (36) \cite{smalley2011}, (37) \cite{smith2012}, (38) \cite{simpson2011}, (39) \cite{wilson2008}, (40) \cite{maxted2015}, (41) \cite{sanchis-Ojeda2011}, (42) \cite{maxted2011}, (43) \cite{southworth2016}, (44) \cite{anderson2012}, (45) \cite{hellier2012}, (46) \cite{sanchis-Ojeda2015}, (47) \cite{triaud2010}, (48) \cite{gillon2011}, (49) \cite{faedi2013}, (50) \cite{hebrard2013}, (51) \cite{mancini2018}, (52) \cite{gillon2013}, (53) \cite{gomez-maqueo-chew2013}, (54) \cite{anderson2014}, (55) \cite{hellier2014}, (56) \cite{wilson2006}}
\tablecomments{The values of $\teff$ were obtained from SWEET-Cat \citep{santos2013}, and the obliquities $\lambda$ from the TEPCat compilation \citep{southworth2011}. The column $P_{\star}^{(\rm phot)}$ indicates the photometric rotation period, while the column $P_\star^{(v \sin i)}$ indicates the value calculated from the $v \sin i$ and stellar radius. The age column indicates the isochronal ages derived in this work, with the values and error bars indicating the 50th, 16th, and 84th percentiles of the posterior distribution.}
\end{deluxetable*}

$ $

\end{document}